\documentclass[preprint,showpacs,preprintnumbers,amsmath,amssymb]{revtex4}
\usepackage{graphicx}
\usepackage{floatflt}
\usepackage{dcolumn}
\usepackage{bm}

\begin{document}

\preprint{APS/123-QED}
\title{ Exploring the extended density dependent Skyrme effective forces 
for normal and isospin-rich  nuclei  to  neutron stars}

\author{B. K. Agrawal$^1$, Shashi K. Dhiman$^{2,3}$ and Raj Kumar$^3$}
\affiliation{
$^1$Saha Institute of Nuclear Physics, Kolkata - 700064, India\\
$^2$School of Physical Sciences,
Jawaharlal Nehru University,
New Delhi 110067 India.\\
$^3$Department of Physics,
H.P. University, Shimla 171 005
India.}

\begin{abstract}
We parameterize the recently  proposed generalized Skyrme effective
force (GSEF) containing extended density dependence.  The parameters
of the GSEF are determined by the fit to several properties of the
normal and isospin-rich nuclei. We also include in our fit a realistic 
equation of state for the pure neutron matter  up to high densities so
that the resulting Skyrme parameters can be suitably  used to model
the neutron star with the ``canonical" mass ($\sim 1.4 M_\odot$). 
For the appropriate comparison we generate a parameter set for
the standard Skyrme effective force (SSEF) using exactly the same set of
the data as employed to determine the parameters of the GSEF.  We find
that the GSEF yields larger values for the neutron skin thickness which
are   closer to the recent predictions based on the isospin diffusion
data.  The Skyrme parameters so obtained are  employed to compute the
strength function for the isoscalar giant monopole, dipole and quadrupole
resonances.  It is found that in the case of GSEF, due to the the
larger value of the nucleon effective mass the values of centroid
energies for  the isoscalar giant resonances are in better agreement
with the corresponding experimental data in comparison to those obtained
using the SSEF. We also present results for some  of the key properties
associated with the neutron star of ``canonical" mass  and for the one with
the maximum mass.

\end{abstract}
 \maketitle

\section{Introduction}
\label{intro_sec}

The density dependent Skyrme type effective nucleon-nucleon interaction
\cite{Bell56} within the Hartree-Fock (HF) approximation has been
one of the most successful and popular  microscopic tools to describe
the ground state properties of the finite nuclei as well as that of
the symmetric nuclear and pure neutron matters. The pioneering work
\cite{Vautherin72} of implementing the Skyrme type effective force
having only  a linear density dependence was carried out to reproduce
the experimental data on the binding energy and charge rms radii.
The linear density dependence gave rise to the value of nuclear matter
incompressibility coefficient $K_\infty $ in the range of $300 - 400$
MeV which is much higher compared to the experimental value $\sim 220$
MeV.  It was proposed in Ref.  \cite{Zamick73} that in order  to obtain
a reasonable value of $K_\infty$ the linear density dependence in the
Skyrme effective force must be modified to $\rho^\alpha$ with $\alpha $
lying in between $\frac {1}{3}$ to $\frac{2}{3}$. Since then, numerous
attempts have been  made to parameterize the standard Skyrme effective
force (SSEF) which contains only a single density dependent term (e.g.,
see Ref.  \cite{Giai81,Bartel82,Tondeur84,Pearson90,Reinhard95,Chabanat98,
Brown98,Bender03,Goriely05,Agrawal05}).  Recently several parameter sets
for the SSEF are obtained by fits to large set of data comprising  the 
binding energy, charge rms radii and  single-particle energies  for
the nuclei ranging from normal to isospin-rich  ones.  Further, the
Skyrme parameters are constrained by demanding that the equation of
state (EOS) for  the pure neutron matter should be reasonable up to
the densities relevant for studying the properties  of neutron star
\cite{Chabanat98,Goriely05,Agrawal05}.

Recently, studies involving the generalized Skyrme effective force (GSEF)
have been  revisited \cite{Duguet05,Cochet04,Cochet04a,Bhattacharyya05}.
Generalization of the  Skyrme effective force  can be realized
by adding several density dependent terms to each of the three,
namely, local, non-local and spin-orbit parts of the SSEF.  The multi
density-dependent terms used in Refs. \cite{Cochet04,Cochet04a} are of
the form $\rho^{\nu/3}$ with $\nu = 1, 2, \text{ and } 3$ connected
to the Fermi momentum ($k_F$) expansion of the Brueckner $G$ matrix.
Similar density dependence is also found in the energy density functional
derived within the chiral perturbation theory containing the contributions
from one and two pion exchange diagrams up to three-loops \cite{Kaiser02}.
In Refs. \cite{Cochet04,Cochet04a} the GSEF containing extended density
dependence in the local part of the Skyrme interaction is used only  to
study the properties of nuclear matter. It is found that the extended
density dependence gives rise to reasonable EOS for the  infinite nuclear
matter for any given asymmetry.  It has been also  demonstrated that
unlike in the case of SSEF the value of $K_\infty$ and the isoscalar
nucleon effective mass $m^*$ can be determined independently using
the GSEF.  Thus, one may expect the GSEF to reproduce concurrently
the experimental values for the $K_\infty$ and the single-particle
energies in nuclei. We would like to mention that some early  attempts
\cite{Liu91,Liu91a,Liu91b,Farine97} to generalize the density dependence
of Skyrme type effective interaction  were significantly different
compared to the one proposed recently \cite{Cochet04,Cochet04a}. In Ref.
\cite{Farine97} the exponents of the density dependent terms are obtained
by a  fit to several nuclear observables.  Moreover, most of these forces
are plagued by an undesirable feature that they give collapse in the
nuclear matter at high densities which leave them unsuitable  for  their
use for the study of neutron star.

In the present paper we determine the parameters of the GSEF 
by  the fit to a set of experimental data for the binding
energies, charge rms
radii, single particle energies and rms radii of valence
neutron orbits. Our data set used in the fit consists of 13
spherical nuclei, namely, $^{16}$O, $^{24}$O,  $^{40}$Ca,
$^{48}$Ca, $^{48}$Ni, $^{56}$Ni, $^{68}$Ni, $^{78}$Ni, $^{88}$Sr,
$^{90}$Zr, $^{100}$Sn, $^{132}$Sn and $^{208}$Pb.
  We also fit the experimental data for
the breathing mode energies for the $^{90}$Zr, and
$^{208}$Pb nuclei. Further, we constrain the value of Skyrme parameters
by including in the fit a realistic  EOS for the pure neutron matter.
The  chi-square minimization, required to obtain the best fit parameters,
is achieved by using the simulated annealing method (SAM)  as recently
implemented \cite{Agrawal05} to determine the  parameters of the SSEF.

The present paper is organized as follows. In Sec. \ref{gsef_sec} we
briefly outline the form of the GSEF and the corresponding energy density
functional adopted in the present work. In this section, we also mention
in short the strategies used to evaluate center of mass corrections
to the binding energy and charge radii.  In Sec. \ref{pgsef_sec} we
briefly describe a procedure for minimization of  the $\chi^2$ function
based on the SAM and present  the set of the experimental data along with
the constraints  used in the  fit to determine the values of the Skyrme
parameters. In the same section we list the values of  the parameter sets for
the GSEF and SSEF.  In Sec. \ref{res_sec} we present our results for the
three different fits carried out in this work.  We also present results
for the isoscalar giant monopole, dipole and quadrupole resonances and
some key properties of the neutron  stars obtained using the newly generated
parameter sets.  Finally, in Sec. \ref{conc} we summarize our main results.

\section{Generalized Skyrme effective force}
\label{gsef_sec}
The GSEF used in Refs. \cite{Cochet04,Cochet04a} can be written as,
\begin{eqnarray}
\label{v12}
&V(\vec{ r}_1,\vec{ r}_2)&= t_{0}\left (1+x_{0} P_{\sigma}\right)\delta(\vec{ r}) \nonumber\\
&&+\frac{1}{2}t_{1}\left (1+x_{1} P_{\sigma}\right ) 
\left[\delta(\vec{ r})\vec{ P}^{\prime 2}+\vec{ P}^{2}\delta(\vec{ r})
\right] \nonumber\\
&&+t_{2}\left (1+x_{2} P_{\sigma}\right )\vec{
P}^{\prime} \cdot  \delta(\vec{ r})\vec{ P} \nonumber\\
&&+\sum_{i}t_{3i}\rho^{\alpha_i}\left (1+x_{3i} P_{\sigma}\right)\delta(\vec{ r})
\nonumber\\
&&+iW_{0} \vec{ \sigma} \cdot \left[\vec{ P}^{\prime} \times
\delta(\vec{ r})\vec{ P}\right ]
\end{eqnarray}
with $i = 1,2,3,...$ and $\alpha_i= i/3 $. In Eq. (\ref{v12}),
$\vec{ r}=\vec{ r}_1 - \vec{ r}_2$, $\vec{ P}=\frac{\vec{
\nabla}_1-\vec{\nabla}_2}{2i}$, $\vec{P}^\prime$ is
complex conjugate of $\vec{P}$ acting on the left and
$\vec{\sigma}=\vec{\sigma}_1+\vec{\sigma}_2$, $P_\sigma=\frac{1}{2}\left
(1+\vec{\sigma}_1\cdot\vec{\sigma}_2\right )$.  The SSEF can be
obtained from Eq. (\ref{v12}) simply by setting $t_{3i} = 0 $ for $i \not=
1$ and $\alpha_1 $ is normally taken to be  less than unity. For instance,
in the SIII force \cite{Beiner75} $\alpha_1$ is unity and for the SLy
forces \cite{Chabanat98} $\alpha_1=1/6$ was used.  It may be noted that
we have considered the extended density dependence only for the local
term in Eq. (\ref{v12}). However, on the same analogy the non-local and
spin-orbit terms can also be extended to have density dependence.

The total energy $E$ of the system is given by, 
\begin{equation}
 E=\int {\cal H}(r)d^3r \end{equation}
where, ${\cal H}(r)$ is  the Skyrme energy density functional corresponding to 
Eq. (\ref{v12}) which under the time-reversal invariance  is given by \cite{Vautherin72, Chabanat98},
\begin{equation}
\label{Hden}
{\cal H} = {\cal K} + {\cal H}_\delta  +{\cal H}_\rho+{\cal H}_{\rm
eff} +{\cal H}_{\rm fin} + {\cal H}_{\rm so} +{\cal H}_{\rm sg}+{\cal
H}_{\rm Coul}
\end{equation}
where, ${\cal K} = \frac{\hbar^2}{2m}\tau$ is the kinetic energy term,
${\cal H}_\delta$ is the zero-range  term, ${\cal H}_\rho$ the density
dependent term, ${\cal H}_{\rm eff}$ an effective-mass term, ${\cal
H}_{\rm fin}$ a finite-range term, ${\cal H}_{\rm so}$ a spin-orbit term,
${\cal H}_{\rm sg}$ a term due to tensor coupling  with spin and gradient
and ${\cal H}_{\rm Coul}$ is the contribution to the energy density for
protons due to the Coulomb interaction. For the Skyrme interaction of
Eq. (\ref{v12}), we have,
\begin{equation}
{\cal H}_\delta = \frac{1}{4}t_0\left [(2 + x_0)\rho^2 - (2x_0 + 1)(\rho_p^2 + \rho_n^2)\right ],
\end{equation}
\begin{equation}
{\cal H}_\rho = \frac{1}{4}\sum_{i }t_{3i}\rho^{\alpha_i}\left [(2 +
x_{3i})\rho^2 - (2x_{3i} + 1)(\rho_p^2 + \rho_n^2)\right ],
\end{equation}
\begin{equation}
{\cal H}_{\rm eff} = \frac{1}{8} \left [t_1 (2 + x_1) + t_2
(2 + x_2)\right ] \tau\rho + \frac{1}{8}\left[ t_2(2x_2+1)
-t_1(2x_1+1)\right] (\tau_p\rho_p + \tau_n\rho_n),
\end{equation}
\begin{eqnarray}
{\cal H}_{\rm fin}&=&\frac{1}{32}\left [3t_1 (2 + x_1) - t_2 (2 + x_2)\right ](\nabla\rho)^2 \nonumber\\
&& - \frac{1}{32}\left[ 3t_1(2x_1+1) +  t_2(2x_2+1) \right] \left [ (\nabla\rho_p)^2 + (\nabla\rho_n)^2\right], \\
{\cal H}_{\rm so}&=&\frac{W_0}{2}\left [ {\bf J}\cdot\nabla\rho +
{\bf J_p}\cdot{\bf \nabla}\rho_p
+{\bf J_n}\cdot{\bf \nabla}\rho_n\right ],\\
\label{Hsg}
{\cal H}_{\rm sg} &=&-\frac{1}{16}(t_1 x_1 + t_2 x_2){\bf J}^2 + 
\frac{1}{16}(t_1 - t_2) \left [{\bf J_p}^2 + {\bf J_n}^2 \right],\\ 
\label{Hcoul}
{\cal H}_{\rm Coul}(r)&=& \frac{1}{2}e^2\rho_p(r)
\int \frac{\rho_p( r^\prime)d^3r^\prime}{\mid \bf r - \bf r^\prime\mid }
-\frac{3}{4}e^2\rho_p(r) \left ( \frac{3 \rho_p(r)}{\pi}\right )^{1/3}.
\end{eqnarray}

Here, $ \rho = \rho_p + \rho_n,$ $\tau = \tau_p + \tau_n,$ {\rm and}
${\bf J}= {\bf J}_p + {\bf J}_n$ are the particle number density, kinetic
energy density and spin density with $p$ and $n$ denoting the protons
and neutrons, respectively.  We have used the value of $\hbar^2/2m =
20.734$ MeV fm$^2$ in our calculations. We would like to emphasize that we
have  included the contributions from the spin-density term as given by
Eq. (\ref{Hsg}) so that the corresponding contributions
can be considered consistently in evaluating the Landau
parameter $G_0^\prime$ \cite{Bender02}. Although
the contributions from Eq.  (\ref{Hsg}) to the binding energy and
charge radii are not very significant, but they may
contribute significantly to the value 
of the Landau parameter $G_0^\prime$.

The single-particle wave functions $\phi _{i}$ and corresponding
single-particle energies $\varepsilon _{i} $ are obtained by solving
the HF equations given by,
\begin{equation}
\left[ -\vec{\nabla }\frac{\hbar ^{2}}{2m_{q}^{\ast }(\mathbf{
r})}\cdot \vec{\nabla }+U_{q}(\mathbf{r})-i\vec{W}_{q
}(\mathbf{r})\cdot\left( \vec{\nabla }\times \vec{\sigma } \right) \right]
\,\phi _{i}(\mathbf{r},q)=\varepsilon _{i}\phi _{i}
( \mathbf{r},q),
\label{HFEq}
\end{equation}
where, $m^*_q$ is the effective nucleon mass, $U_q$ and $W_q$
are the central and spin-orbit parts of the mean field potentials.
The expressions for the $m^*_q$ and $W_q$
can be found  in  Ref.  \cite{Chabanat98}. The
expression for $U_q$ is given as,
\begin{eqnarray}
U_{q }(\mathbf{r}) &=&\frac{1}{2}t_0\left[ (2+x_0)\rho ({\mathbf
r})-( 1+2x_0)\rho _{q }(\mathbf{r}) \right ]\nonumber \\
&&+\frac{1}{8}\left[
t_1(2+x_1)+t_2(2+x_20)
\right] \tau ({\mathbf r})-\frac{1}{8}\left[
t_1(1+2x_1)-t_2(
1+2x_2)\right] \tau _{q }({\mathbf r})  \nonumber \\
&&+\frac{1}{4}\sum_{i}t_{3i}\left\{(2+x_{3})(2+\alpha_i)\rho ^{\alpha_i+1}
({\mathbf r})-(1+2x_{3})\left [\alpha_i \rho ^{\alpha_i -1}({\mathbf r}
)(\rho _{p }^{2}(\mathbf{r})+\rho _{n}^{2}(\mathbf{r}))
+2\rho ^{\alpha_i}({\mathbf r})\rho _{q}\right]\right \}
\nonumber \\
&&-\frac{1}{16}\left[
3t_1(2+x_1)-t_2(2+x_2)
\right] \nabla ^{2}\rho ({\mathbf r})  \nonumber \\
&&+\frac{1}{16}\left[
3t_1(1+2x_1)+t_2(1+2x_2)
\right] \nabla ^{2}\rho _{q }({\mathbf r})\nonumber \\
&& -\frac{1}{2}W_0\left (\vec{\nabla}\cdot
\vec{J}+\vec{\nabla}\cdot
\vec{J}_q\right )
 +\delta _{q ,p}\, V_{\rm Coul}(\mathbf{r}),
\label{U-tau}
\end{eqnarray}
where, $q = p, n$.

To this end we briefly outline the corrections made to the HF results for
the binding energies and charge rms radii arising from various effects
as follows.  The center of mass (CM) corrections to the binding energies
and charge rms radii are made using simple prescriptions discussed in
Ref. \cite{Shlomo78,Agrawal05}.  We also consider the corrections to
the charge rms radii due to the spin-orbit effect as well as the  charge
distributions of the neutron and proton \cite{Bertozzi72,Chabanat98}. The
contribution from the Coulomb exchange term in Eq.  (\ref{Hcoul})
is dropped in order to compensate for the  effects of the long-range
correlations \cite{Shlomo82} on the Coulomb displacement energy (CDE)
in mirror nuclei. We also add the modified Wigner term \cite{Goriely02},
\begin{equation}
E_w=V_w \text {exp}\left\{-\lambda\left (\frac{N-Z}{A}\right )^2\right\}
+V_{w^\prime}\left | N-Z\right |\text {exp}\left\{-\left
(\frac{A}{A_0}\right)^2\right \}
\label{wig-corr} \end{equation} 
to the binding energy. Values of the parameters $V_w$, $V_{w^\prime}$,
$\lambda$ and $A_0$ are determined by the fit.

\section{Parameterization of the generalized Skyrme effective force}
\label{pgsef_sec}
The parameters of the GSEF can be determined by 
minimizing the value of $\chi^2$ which is given
as, \begin{equation} \chi^2 =  \frac{1}{N_d - N_p}\sum_{i=1}^{N_d}
\left (\frac{ M_i^{exp} - M_i^{th}}{\sigma_i}\right )^2 \label {chi2}
\end{equation}
where, $N_d$ is the number of  experimental data points and $N_p$ the
number of fitted  parameters. The $\sigma_i$ stands for theoretical error
and $M_i^{exp}$ and $M_i^{th}$ are the experimental and the corresponding
theoretical values, respectively, for a given observable.  The values
of $\chi^2 $ depends on the Skyrme parameters, since, the $M_i^{th}$
in Eq. (\ref{chi2})  is calculated using the HF approach with a  Skyrme
type effective nucleon-nucleon  interaction.  Thus, it is clear that
the best fit parameters can be obtained using an appropriate method for
the $\chi^2$ minimization together with a set of experimental data. We
briefly describe below the SAM for the $\chi^2$ minimization and the
set of experimental data selected to obtain the best fit parameters.

\subsection{SAM algorithm for $\chi^2$ minimization}
\label{fit_pro}
The concept of SAM is based on the manner in which liquids freeze or
metals recrystallize in the process of annealing. In an annealing process
a metal, initially at high temperature and disordered, slowly cools
so that the system at any time is in a  thermodynamic equilibrium. As
cooling proceeds, the system becomes more ordered and approaches a frozen
ground state at zero temperature.  The SAM is an elegant technique  for
optimization problems of large scale, in particular, where a desired
global extremum is hidden among many local extrema.  This method has been
found to be an extremely useful tool for a wide variety of minimization
problems of large non-linear systems in many different areas of science
(e.g., see Refs. \cite{Patrik84,Ingber89,Cohen94}).  Very recently
\cite{Burvenich02,Burvenich04}, the SAM was used to generate some initial
trial parameter sets for the  point coupling variant of the relativistic
mean field model.

In Ref. \cite{Agrawal05} we have implemented the SAM to  the problem of
searching the global minimum in the hypersurface of $\chi^2$ function
as defined by Eq. (\ref{chi2}). Here too we shall use the SAM to
determine the parameters of the GSEF.  In the SAM one needs to specify
the appropriate annealing schedule together with the parameter space
(i.e., the range of the values of the parameters) in which the best fit
parameters are to be searched.  Similar to that  in Ref. \cite{Agrawal05},
in the present work we have employed a moderately faster annealing
schedule such as Cauchy annealing schedule given by, \begin{equation}
T(k)=T_i/k \label{Tk} \end{equation} where, $T_i$ is the initial value
of the control parameter (also viewed as effective temperature) and $T(k)$
with $k= 1, 2, 3,...$ is the control parameter at $k-$th step. The value
of $k$ is increased by unity after $100N_p$  reconfigurations or $10N_p$
successful reconfigurations whichever occurs first.  The value of $T_i$
is taken to be 1.25 which is same as that used in Ref. \cite{Agrawal05}. We
keep on reducing the value of the control parameter using Eq. (\ref{Tk})
in the subsequent steps  until the effort to reduce the value of $\chi^2$
further becomes sufficiently discouraging.  In Ref. \cite{Agrawal05}
instead of the range of the values of the Skyrme parameters we used
range of the values of the quantities associated with the symmetric
nuclear matter.   Since, it was possible to identify the number of nuclear
matter quantities linearly related to the equal number of parameters for
the  SSEF. But, in the case of GSEF the numbers of parameters are larger
and we shall define the parameter space directly in terms of the range of
the values each of the Skyrme parameter can take. In Table \ref{range_skm}
we give the lower and upper limits for the values of the Skyrme parameters
denoted by ${\bf v_0}$ and ${\bf v_1}$, respectively. The quantity ${\bf
d}$ in the penultimate column denotes the maximum displacement allowed
in a single step for a given Skyrme parameter during the reconfiguration
process (see also Ref. \cite{Agrawal05} for detail). The last column
labeled as ${\bf v}_{in}$ contains initial values for the Skyrme
parameters used as a starting point for the SAM.  These values for the
parameters are the same as that for SGI force \cite{Giai81} which has
$\rho^{1/3}$ density dependence. The parameters of the Wigner term (Eq.
\ref{wig-corr}) are taken from Ref. \cite{Goriely02}.  The lower and
upper limits and the initial values for the parameters $\alpha_i$ are
taken to be the same indicating that these parameters are kept fixed
during the $\chi^2$ minimization.

\subsection{Data used in the  fitting procedure}
\label{bain}
We now summarize our selection of data and corresponding theoretical
errors adopted in the fitting procedure.  In Table \ref{in2} we
list our choice of the data along with their sources \cite{Audi03,
Otten89,Vries87,Kalantar88,Platchkov88,Bohr69,BrownP,Youngblood99,Wiringa88}.  It
must be noted that in addition to the data on the binding energy, charge
radii and single particle energies, the values of the Skyrme parameters
are  further constrained   by including in the fit the experimental data
for the radii of valence neutron orbits, breathing mode energies together
with a realistic EOS for the pure neutron matter upto the
densities ($\sim 0.8$ fm$^{-3}$) relevant for the study of neutron stars.
For the binding energy we use in our fit the error of 1.0 MeV except for
the $^{100}$Sn nucleus. The binding energy for the $^{100}$Sn nucleus is
determined from systematics and are expected to have large errors. Thus,
we assign it a  theoretical  error of 2.0 MeV. For the charge rms radii
we use the theoretical error of 0.02 fm except for the case of $^{56}$Ni
nucleus. The charge rms radius for the $^{56}$Ni nucleus is obtained from
systematics and we use the theoretical error of 0.04 fm.  For the rms
radii of the valence neutron orbits in $^{17}$O and $^{41}$Ca nuclei
we use $r_v(\nu 1d_{5/2})=3.36$ fm and $r_v(\nu 1f_{7/2}) = 3.99$ fm,
\cite{Kalantar88,Platchkov88} respectively. The theoretical error taken
for the rms radii for the valence neutron orbits is 0.06 fm. We must point
out that the choice of the theoretical error on the rms radii for the
valence neutron orbits is due to the large uncertainties associated
with their extraction from the experimental measurements. To be
consistent with the way these valence neutron radii are determined, we
do not include the center of mass correction to these data.  For each
of  the  22 single-particle (S-P) energies in the $^{208}$Pb nucleus,
we have used the theoretical error of 1.0  MeV in our fit.  The
experimental data for the breathing mode constrained energies $E_0$
included in our fit are 17.81, and 14.18 MeV  for the  $^{90}$Zr  and
$^{208}$Pb nuclei \cite{Youngblood99}, respectively, with the theoretical
error taken to be 0.5 MeV for the $^{90}$Zr nucleus and 0.3 MeV for the
$^{208}$Pb nucleus.  We also include in the fit the EOS for the pure
neutron matter of a realistic UV14+UVII model \cite{Wiringa88}. We
use 15 data points for the EOS  corresponding to the densities up to
$5\rho_0$. The theoretical error on each of these 15 data points are
taken to be 2.0 MeV.

\subsection{Parameters for the generalized Skyrme effective force}
\label{gsef_par}
Following the fitting procedure described in Sec. \ref{fit_pro} together
with the list of data  given in Table \ref{in2} we have generated two
different parameter sets for the GSEF. These parameter sets
are given in Table \ref{newpar}.  The parameter set GSkI and GSkII
are obtained as follows; GSkI includes the density dependent terms
proportional to $\rho^{\alpha_i}$ with $\alpha_i = 1/3, 2/3$ and 1,
whereas, GSkII force has only $\rho^{1/3}$ and $\rho^{2/3}$ density
dependent terms.
The exponents for the density dependence are taken to be 1/3 and 2/3 for
the GSkII interaction due to the following reasons.
 The $\rho^{2/3}$ density 
 dependence decouples  the effective mass and the nuclear matter incompressibility coefficient \cite{Cochet04}. On the otherhand,  the SSEF  favours the exponent of density dependence to lie in between $1/3- 1/6$ in order to give an acceptable value for the nuclear matter incompressibility coefficient.
  In the last column of Table \ref{newpar} we give the
values of parameters corresponding to the  SSEF \cite{Chabanat98}.  We name
this force as SSk  which is obtained using the same set of data as
those for the GSkI and  GSkII forces.  We shall see in the following
section that the quality of the fit for the GSkI force is better in
comparison to the ones obtained using GSkII and SSk forces.  It can be
seen from Eq. (\ref{v12})  that  the density dependence in the GSkI
force requires six parameters, namely, $t_{3i}$ and $x_{3i}$ with $i = 1, 2$ and
$3$, whereas, in the SSk force  the density dependence is specified by
three parameters, $t_{31}$, $x_{31}$ and $\alpha_1$. Thus, 
fit to the GSEF considered here  requires three additional parameters.

\section{Results and discussions}
\label{res_sec}

We  generate the parameter sets  GSkI and  GSkII for the GSEF using
the experimental data on the bulk properties for the nuclei
ranging from the normal to isospin-rich  ones. The values for the Skyrme
parameters are further constrained by including in fit the experimental
data on the breathing mode energies  and a realistic EOS for the pure
neutron matter. For the appropriate  comparison we also obtain a parameter
set  named SSk for the  SSEF  using the same set of data as used to
determine the parameters of the GSEF.  The values of parameters for all
the three forces in consideration are given in Table \ref{newpar}. In
the following subsections we shall present the results obtained using
these three different parameter sets.

\subsection{Infinite symmetric and asymmetric nuclear matter}

In Table \ref{nm_par} we give the values of various quantities associated
with the symmetric nuclear matter at the saturation density obtained
for GSkI, GSkII and SSk forces.  We see that the values for all the
quantities except for the
$m^*$ and
\begin{equation}
L = 3\rho \frac{dJ}{d\rho},
\label{L}
\end{equation}
given in Table \ref{nm_par} are quite close to each other
for all the three forces.  The values of $m^*$ and $L$ for the GSkI
and GSkII forces are little larger than that for the SSk force. The value
of $m^*\sim 0.8m$ for the GSkI and  GSkII forces is highly desirable
in order to appropriately reproduce the location of the isoscalar giant
quadrupole resonance \cite{Reinhard99}. Further, we shall see below that
larger value of $m^*$ gives rise to the single-particle energies which
are in better agreement with the experimental data.  The values of $L$
at the saturation density obtained for the GSkI and GSkII  forces in
comparison to the one for the SSk force are in better agreement with
the recently extracted value of $L = 88\pm 25$ MeV \cite{Chen05}.

We consider  the behavior of  the symmetry energy coefficient $J(\rho)$
for densities relevant to the study of neutron stars. It is well known
\cite{Kutschera94,Kutschera00} that the values of $J(\rho)$ and the
resulting EOS for the pure neutron matter at higher densities ($\rho >
2\rho_0$, $\rho_0 =0.16 $ fm$^{-3}$) are crucial in understanding the
various properties of neutron star. For example, the proton fraction at
any density depends strongly on the value of $J(\rho)$ at that density,
which in turn affects the chemical compositions as well as the cooling
mechanism of the neutron star \cite{Lattimer91}.  Yet, no  consensus
is reached for the density dependence of $J(\rho)$.  We display in
Fig. \ref{S_rho}, our results for the variation of the symmetry energy
$J$ as a function of the nuclear matter density $\rho$. We see that
the  value of $J$ increases with density.  The slope of $J(\rho)$ being
positive  upto $3\rho_0$ for GSkI, GSkII and SSk forces indicate that
these interactions can be used to study the properties of the neutron
star with masses around $1.4M_\odot$ \cite{Stone03}. In Fig. \ref{eos_nm}
we plot the EOS for the nuclear matter for various proton fraction
( $Y_p=\rho_p/\rho$) as a function of density.  The solid squares
represent the EOS for the  pure neutron matter (i.e.,$Y_p=0$) for a
realistic UV14+UVII model \cite{Wiringa88}. It can be seen that the
EOS for the pure neutron matter obtained using GSkI force is in better
agreement with the one obtained for the UV14+UVII model.

\subsection{Fit to the nuclear bulk properties}

In Fig. \ref{del_be}  we present our results obtained using the
GSkI, GSkII and SSk forces for the relatives errors in the
values of total binding energies 
 for the nuclei used in the fit.  
The rms error for the total binding energy are 1.18, 1.72 and 1.28 MeV
for the GSkI, GSkII and SSk forces, respectively.
Thus, We can say that the quality of the fit to the binding
energies for the GSkI and SSk forces are more or less the
same and they are much better than that for GSkII force.
The binding energy difference $B(^{48}{\rm Ca}) - B(^{48}{\rm Ni})$
= 66.82, 66.57 and  66.95 MeV for the GSkI, GSkII and SSk interactions,
respectively, compared to the experimental value of 68.85 MeV.  The said
difference for the SKX interaction \cite{Brown98}  is 66.3 MeV which is
about 0.5  MeV lower  as compared to our present results.  On the other
hand,  most of the Skyrme interactions which include the contribution
from the exchange Coulomb term yield $B(^{48}{\rm Ca})-B(^{48}{\rm
Ni})\approx 63$ MeV, which is about 6 MeV lower than the corresponding
experimental value. 
To this end we would like to remark that the Wigner corrections (Eq.
(\ref{wig-corr})) are more significant for the GSk forces as compared to the one
for the SSk force. It may be verified by using  the values of the Wigner
parameters  given in Table
\ref{newpar}
that for the symmetric nuclei the Wigner correction is greater than 2.5 MeV
for the GSk forces which is more than twice than that for the SSk force.
Also for asymmetric nuclei  we find that the GSk forces has larger Wigner
corrections for the ligher nuclei. For example, in case of $^{24}$O the Wigner correction is about 4.0(0.5) MeV for the GSk(SSk) forces.  
In Fig. \ref{del_rc} we present our
results  for the
relative errors in the values of charge rms radii for the
nuclei used in the fit.
The rms error for the charge rms radii are 0.023, 0.026 and 0.021 fm
for the GSkI, GSkII and SSk forces, respectively.
  The quality of fit for the charge rms radii is more or less
same  for all the three forces considered here.

In Fig. \ref{pb208_spe} we give the values for the single-particle
energies for the $^{208}$Pb nucleus used in the fits. We  see that the
values of the single-particle energies for the case of GSkI and GSkII
forces are in slightly better agreement with the experimental data as
compared to the ones obtained for the SSk force. This is mainly due
to the weakening of the correlations between the effective mass $m^*$
and the incompressibility coefficient $K_\infty$ in the case of the
GSEF \cite{Cochet04,Cochet04a}.  In Table \ref{extra_data} we present our
results for the rms radii for the valence neutron orbits and the breathing
mode constrained energies used in the fit. The breathing mode constrained
energy $E_0$ is given by, \begin{equation} E_0 = \sqrt{\frac{m_1}{m_{-1}}}
\label{econ} \end{equation} where, fully self-consistent values for the
energy-weighted ($m_1$) and inverse energy-weighted ($m_{-1}$) moments
of the strength function calculated using random phase approximation
(RPA)  for the breathing mode (or isoscalar giant monopole resonance)
can be obtained using the double commutator sum rule and the constrained
HF method \cite{Bohigas79,Agrawal05}, respectively.  We see that all
the Skyrme forces under consideration gives the similar fit to the
experimental data presented in Table \ref{extra_data}.

\subsection{Neutron skin thickness}
  The neutron skin thickness $r_{n}-r_{p}$  is  the
difference between the rms radii obtained using the density distributions
for the point neutrons and point protons. 
It is well known \cite{Brown00a,Horowitz01}  that the accurate measurement
of the neutron skin thickness  would place a stringent constraint on
the density dependence of the nuclear symmetry energy. In particular,
it has been shown \cite{Chen05} that the neutron skin thickness in
heavy nuclei are linearly correlated with the slope of the symmetry
energy coefficient $L$. The value of $L$ is poorly known  due to the
large uncertainty associated with the measurement of the neutron skin
thickness. The measured values of the neutron skin thickness is very
much probe dependent.  Current data indicate that the neutron skin
thickness in the $^{208}$Pb nucleus lie in the range of $0.10 - 0.28$
fm \cite{Krasznahorkay04}.  The proposed experiment at the Jefferson
Laboratory on parity violating electron scattering from $^{208}$Pb is
expected to give another independent and more accurate measurement (within
0.05 fm) of its neutron skin thickness. Recent analysis of the isospin
diffusion data \cite{Chen05} yields $L = 88 \pm 25$ MeV which restricts
the values of the neutron skin in the $^{132}$Sn and $^{208}$Pb nuclei
to $0.29 \pm 0.04$ fm and $0.22 \pm 0.04$ fm,
respectively. In Fig. \ref{nskin}
 we present our results for neutron skin thickness obtained using GSkI,
GSkII and SSk forces.  The values for neutron skin
thickness for the GSkI and GSkII
forces are larger than those for the SSk force. It is because of the
fact that  the values of $L$ for the GSkI and GSkII forces are larger
than that for the SSk force (see Table \ref{nm_par}).  The
values of neuron skin thickness obtained using the GSk forces for the $^{132}$Sn
and $^{208}$Pb nuclei are in better agreement with the recent predictions
$0.29 \pm 0.04$ and $0.22 \pm 0.04$, respectively.

\subsection{Isoscalar giant resonances}
We consider now the results for the isoscalar giant monopole, dipole
and quadrupole resonances obtained using the GSkI, GSkII and SSk
forces. The centroid energy for the isoscalar giant
monopole resonance (ISGMR) depends mainly on the value of $K_\infty$
\cite{Blaizot95}. The centroid energy for the overtone mode of  the
isoscalar giant dipole resonance (ISGDR)  is governed by the values
of the $m^*$ as well as $K_\infty$ \cite{Kolomietz00}, whereas, the
centroid energy of the isoscalar giant quadrupole resonance (ISGQR)
being a surface oscillation depends merely on the value of $m^*$
\cite{Reinhard99}.  As we mentioned earlier, the correlations
between the $m^*$ and $K_\infty$ is weaker for the GSEF which may give
rise to a better agreement of the experimental data for the isoscalar
giant resonances.

The excitation energy dependence of the strength function $S(E)$  for
the isoscalar giant resonances is obtained using the 
RPA Green's function $G$ as \cite{Shlomo75},
\begin{equation}
S(E)\!=\!\sum\limits_{n}\left|\langle 0 |F| n\rangle\right|^2
\delta(E-E_n)=\frac{1}{\pi}\mbox{Im}\left[\mbox{Tr}(fGf)\right],
\label{strace_eq}
\end{equation}
where,
\begin{equation}
F=\sum\limits_{i=1}^{A} \hat{f}({\bf r}_i)\,
\label{equ:scop}
\end{equation}

is the scattering operator with 
\begin{equation}
\label{f_isgr}
\hat {f}({\bf r})= \left\{
\begin{array}{c}
r^2Y_{00}(\hat{\bf r})\qquad\qquad\qquad\,\,\,\,\,\text{ISGMR}\\
(r^3-\eta r)Y_{10}(\hat{\bf r})\qquad\qquad\text{ISGDR}\\
r^2Y_{20}(\hat{\bf r})\qquad\qquad\qquad\,\,\,\,\,\text{ISGQR}\\
\end{array}
\right .
\end{equation}
for the various isoscalar giant resonances considered. The quantity $\eta$
in Eq. (\ref{f_isgr}) for the case of ISGDR is taken to be equal to
$\frac{5}{3}\langle r^2\rangle$ in order to eliminate the spurious
contributions arising from the center of mass motion.  The RPA
Green's function appearing in the Eq. (\ref{strace_eq}) is obtained by
discretizing the continuum \cite{Shlomo02}. The continuum is discretized
by a box of 30 fm with the radial mesh size of 0.3 fm. The strength
functions are smeared using a Lorentzian with the smearing parameter taken
to be equal to 1.0 MeV for the purpose of plotting.  We must mention that
the contributions due to the spin-orbit and Coulomb interactions are
ignored in evaluating the RPA Green's function, but, they are present
in the HF calculations. However, neglect of spin-orbit and Coulomb
interactions do not have any important bearing on our findings. It has
been shown in Refs.  \cite{Agrawal04,Tapas05} that ignoring the spin-orbit
and Coulomb interactions in the RPA calculations do not affect the values
of the centroid energies for $^{90}$Zr and $^{208}$Pb nuclei as 
the effects of ignoring the spin-orbit interaction is counterbalanced
by that for the Coulomb interaction. Further, in the present work we
are mainly interested in the differences between the behaviour of the
isoscalar giant resonances for the different interactions.

In Figs. \ref{isgmr}a - \ref{isgmr}c we have plotted the strength function for the ISGMR,
ISGDR and ISGQR for the $^{208}$Pb nucleus obtained using GSkI, GSkII
and SSk forces. We see that for the case of ISGMR all the three forces
give similar results. But, for the case of ISGDR and ISGQR the peaks
at the  high energy for the GSkI and GSkII forces are lowered by about
0.5 MeV in comparison to those for the SSk force. Similar
trends are also observed in the $^{90}$Zr nucleus (not shown).  In Table
\ref{ecen} we present our results for the centroid energy $\left
(\frac{m_1}{m_0}\right )$, constrained energy $\sqrt{\frac{m_1}{m_{-1}}}$
and scaling energy $\sqrt{\frac{m_3}{m_1}}$ for the ISGMR, ISGDR and ISGQR
for the $^{90}$Zr and $^{208}$Pb nuclei obtained using GSkI, GSkII and
SSk forces.  
The values of the $k$-th moment $m_k$ are obtained using the strength
function (Eq. \ref{strace_eq}) as,
\begin{equation}
m_k = \int_{E_1}^{E_2} E^k S(E) dE
\label{mk}
\end{equation}
where, $E_1$ and $E_2$ are the minimum and maximum values of the
excitation energy over which the integration is performed.  The range
of the excitation energy $E_1 - E_2$ used to obtain various energies
for the isoscalar giant resonances presented in Table \ref{ecen}
are as follows.  To evaluate the ISGMR energies we have used $E_1 -
E_2$ to be  $0 - 40$ MeV for all the cases considered. For obtaining
the values of ISGDR and ISGQR energies  for the $^{90}$Zr ($^{208}$Pb)
nuclei we use the integration range for the excitation energy as $18 -
40 (16 - 40)$ and $9.5 - 40 (7.5 - 40)$ MeV, respectively.  The non-zero
values for the $E_1$ for the case of ISGDR and ISGQR are so choosen that the
low-lyings peaks are excluded from the calculation of the $k-$th moments
given by Eq. (\ref{mk}).  The strength functions used in Eq. (\ref{mk})
to evaluate the moments $m_k$ are smeared using a small value of the
smearing parameter  $\Gamma/2  = 0.05 $ MeV.  It is clear from this
table that the centroid energies for the ISGDR and ISGQR obtained using
GSkI and GSkII forces are in better agreement with the corresponding
experimental data in comparison to those obtained using SSk force.

\subsection{Properties of the neutron star}

In this section we present our results mainly for the non-rotating
neutron star obtained using the newly generated Skyrme parameter sets.
In particular,   we have studied the relationship between the mass and
radius for the non-rotating neutron star and obtained some of the key
properties for the neutron star with the ``canonical" mass of $1.4M_\odot$
and for the one with maximum  mass $M_{max}$.  Bulk properties of  spherically
symmetric non-rotating neutron star can be obtained by integrating  the
Tolman-Oppenheimer-Volkoff (TOV) equation given as,

\begin{equation}
\frac{dP}{dr}=-\frac{G \varepsilon(r){ M}(r)}{r^2 c^2}
\left [1+\frac{P(r)}{\varepsilon(r)}\right ]
\left [1+\frac{4\pi r^3P(r)}{{ M}(r)c^2}\right ]
\left [1-\frac{2 G { M}(r)}{r c^2}\right ]^{-1},
\label{toveq1}
\end{equation}
and
\begin{equation}
\frac{d{ M}(r)}{dr}=\frac{4\pi r^2\varepsilon(r)}{c^2},
\label{toveq2}
\end{equation}
where, the quantities $P$, $\varepsilon$ and ${ M}$ are the
pressure, energy density and the gravitational mass of the neutron
star, respectively, which depend on the radial distance $r$ from
the center. Numerical solutions of the Eqs.  (\ref{toveq1}) and
(\ref{toveq2}) can be easily obtained for given initial values for
the pressure $P(0)$ at the center and using ${ M}(0) = 0$, provided,
the relation between $P$ and $\varepsilon$ is known.  For lower
densities, we use Baym-Pethick-Sutherland EOS \cite{Baym71}, matching
onto the Skyrme EOS at $\rho \sim 0.5\rho_0$ and going down to $\rho =
6.0\times 10^{-12}$ fm$^{-3}$.  At densities larger than $0.5\rho_0$ we
use the Skyrme EOS which is obtained by adding the nuclear and leptonic
contributions .  The nuclear part of the energy density is obtained
using Eq. (\ref{Hden}) and following the Ref.  \cite{Chabanat97}. In the
leptonic sector we have considered the contributions from the electrons
and muons with the corresponding energy densities $\varepsilon_e$ and
$\varepsilon_\mu$ given by,
\begin{equation}
\varepsilon_l=\frac{1}{\pi^2}\int_0^{k_f^l}k^2\sqrt{k^2+m_l^2} dk,
\end{equation}
where, $l = e$ (electron) or $\mu$ (muon) in the above equation and
$k_f^l$ is the corresponding fermi momentum.  For a given baryon density
$\rho$, the values of the fermi momenta for the neutrons, protons, electrons
and muons can be obtained by ensuring that the conditions for charge
neutrality and chemical equilibrium are satisfied, i.e.,
\begin{eqnarray}
\rho_p&=&\rho_e+\rho_\mu, \\
\label{chemeq1}
\mu_n&=&\mu_p+\mu_e,\\
\label{chemeq2}
\mu_e&=&\mu_\mu.
\end{eqnarray}
In Eqs. (\ref{chemeq1}) and (\ref{chemeq2}), $\mu_j$
denotes  the chemical potential with $j = n, p, e$ or $\mu$.
Thus, the energy density depends only on the baryon density
and the pressure can be obtained as,
\begin{equation}
P=\rho^2\frac{d(\varepsilon/\rho)}{d\rho}\, .
\end{equation}

We now consider our  results for the properties of neutron stars for the
GSkI and SSk forces.  Results for the neutron star with the canonical
mass of $1.4M_\odot$ obtained using the GSkI and GSkII forces are very
much similar.  Because, for both of these forces the density dependence
of the symmetry energy and the pure neutron matter are quite close up
to the densities $3\rho_0$. However, the GSkII force is not appropriate
for the study of the neutron stars with masses larger than the canonical
ones due to the undesirable feature for the symmetry energy at densities
higher than $3\rho_0$ (see Fig.  \ref{S_rho}).  In Fig. \ref{Yj} we
have plotted the variations of the proton fraction ($Y_p=\rho_p/\rho$)
and electron fraction ($Y_e = \rho_e/\rho$) as a function of baryon
density.  The values for $Y_p$ and $Y_e$ or $Y_\mu$ at all the densities
considered are not large enough to allow the direct Urca process.
In Fig. \ref{mrho} we display our results for the variations of the
neutron star mass as a function of the  central density $\rho_c$ for the
baryons. The relationship between the mass of the neutron star and its
radius is shown in Fig. \ref{mrns}.  In Table \ref{ns_prop} we present
our results for some of the key properties associated with the neutron
star with masses $1.4M_\odot$ and $M_{\rm max}$. These properties are
quite similar to the ones obtained using SLy4 interaction \cite{Stone03}.
The quantity $\varepsilon_c/c^2$ is the energy density at the center.
The values for baryon number ($A$) and binding energy ($E_{\rm bind}$)
for neutron star given in  Table \ref{ns_prop} are obtained as follows,
\begin{equation} A=\int_0^R \frac{4\pi r^{2} \rho(r) dr }{\left
(1-\frac{2GM}{rc^2}\right )^{1/2}}, \label{bnum} \end{equation} and
\begin{equation} E_{\rm bind}=(Am_0-M)c^2 \label{ebind} \end{equation}
where, $m_0= 1.66\times 10^{-24}$g is the mass per baryon in $^{56}$Fe.
We also give in Table \ref{ns_prop} the value of  gravitational redshift
of photons ($Z_{\rm surf}$) emitted from the neutron star  surface which
is given  as, \begin{equation}
 Z_{\rm surf}=\left (1-\frac{2GM}{Rc^2}\right )^{-1/2} - 1.
\label{zsurf} \end{equation} 
In Eqs. (\ref{bnum}) - (\ref{zsurf}), $R$ is the radius of neutron star
with the gravitational mass $M$.  We must point out that softening of
the EOS at higher densities due to appearance of the hyperons and kaons
might alter the results.  In particular, results for the neutron star
with the maximum  mass is expected to get affected  when contributions
from the hyperons and kaons are also included in the EOS.

In Fig. \ref{rp4} we have plotted the variation of $R P^{-1/4}$   as
a function of the baryon density. Here, $R$ is the radius (in km)
of the neutron star with $1.4 M_\odot$ and $P$ the pressure in units
of MeV/fm$^3$. The solid circles denote the empirical values for the
$RP^{-1/4}$ at the baryon densities $\rho/\rho_0 = 1.0, 1.5$ and 2.0
taken from Ref.  \cite{Lattimer05}.  In Fig. \ref{ebind_beta} we plot
the ratio of neutron star binding energy to its gravitational  mass
as a function of the compactness parameter $\beta = \frac{GM}{Rc^2}$,
$R$ being the radius of the neutron star with gravitational mass $M$. The  solid
circles represent the empirical values obtained using the relation
\cite{Lattimer01}, 
\begin{equation} \label{ebind1} \frac{E_{\rm
bind}}{Mc^2}=\frac{0.6\beta}{1 - 0.5\beta}.  \end{equation} 
Finally, in Fig. \ref{mi} we plot the variation of moment of inertia
$I$ versus the neutron star mass. To obtain the values of $I$ we have
used the code written by Stergioulas \cite{Stergioulas95}. The results
presented in Fig. \ref{mi} are obtained at the Keplerian frequencies.

\section{Conclusions}
\label{conc}
We have parameterized the GSEF containing multi density dependent terms
of the form $\rho^\nu$ with $\nu = 1/3, 2/3$ and 1 in the local part
of the Skyrme interaction. The experimental data for the normal and
isospin-rich nuclei are used to fit the parameters of the GSEF.  Further,
a realistic equation of state for the pure  neutron matter upto high
densities ($\sim 0.8$ fm$^{-3}$) is used in the fit to ensure that the Skyrme
parameters so obtained can be used to study the neutron star properties.
For the appropriate comparison we generate a parameter set
for the SSEF using exactly the same set of data as in the case of
the GSEF. Comparing our results for the various quantities associated
with the symmetric nuclear matter (at the saturation density) obtained
using the  parameters of GSEF and SSEF we find that the earlier one
yields larger values for the isoscalar effective nucleon mass and
for the quantity $L=3\rho\frac{dJ}{d\rho}$ which is directly related
to the slope of the symmetry energy coefficient $J$. The large
value of the isoscalar effective nucleon mass is highly desirable in order to
predict appropriately the location of the isoscalar giant quadrupole
resonance. The value of $L$ obtained for the GSEF 
is in better agreement with the ones extracted very recently \cite{Chen05}
from the isospin diffusion data. The large value of $L$ obtained for  the
parameters of the GSEF gives rise to the larger values for the neutron
skin thickness corroborating recent predictions \cite{Chen05,Todd-Rutel05}.

The parameter sets generated in the present work for the GSEF and
SSEF are used to calculate the strength function for the isoscalar
giant resonances and some of the  key properties for non-rotating
neutron star. Calculations are performed for the strength function
of the isoscalar giant monopole, dipole and quadrupole resonances for
the $^{90}$Zr and $^{208}$Pb nuclei. We find that the results for the
isoscalar giant monopole resonance are quite similar for the GSEF and
SSEF parameters. However, the larger isoscalar effective nucleon mass
in the case of GSEF  lowers the energy of the isoscalar giant dipole
and quadrupole resonances by $\sim 0.5$ MeV and thereby improve the
agreement with experimental data. The results for the mass-radius
relationship and some of the  key properties for the neutron star with
the ``canonical" mass of $1.4 M_\odot$ and for the one with the maximum
mass are more or less same for GSEF and SSEF. We have also
studied the variation of the $RP^{-1/4}$  (see Fig. \ref{rp4}) as a
function of the baryon density which reasonably agrees with the empirical
values \cite{Lattimer05}. The comparison with the empirical values were
also done for the variation of the ratio of the binding energy of the
neutron star to its gravitational mass as a function of the compactness
parameter.

\begin{acknowledgments}
We would like to thank J. N. De and  S. K. Samaddar for their helpful
comments  and for reading the manuscript. We also acknowledge  the
discussions with D. Bandyopadhyay.

\end{acknowledgments}

\newpage

\newpage
\begin{figure}
\caption{\label{S_rho}
The density dependence of the symmetry energy coefficient $J(\rho)$ 
for the  GSkI, GSkII and SSk forces.}

\caption{\label{eos_nm}
The EOS of nuclear matter for the proton fractions $Y_p = 0,
0.25, 0.4$ and $0.5$ for the GSkI, GSkII and SSk forces. The solid
squares represent the EOS for pure neutron matter (i.e., $Y_p = 0$)
obtained for the UV14$+$UVII model \cite{Wiringa88}.  }

\caption{\label{del_be}
Comparison of the results for the relative errors,
$\Delta B= (B^{expt}-B^{th})/B^{expt}$, in the values of binding energies for the
nuclei considered in the fit to obtain the
 GSkI, GSkII and SSk forces. A=48 represents the result for $^{48}Ca$
nucleus. The result for $^{48}Ni$ nucleus is discussed in the text.
The rms error for the total binding energy are 1.18, 1.72 and 1.28 MeV
for the GSkI, GSkII and SSk forces, respectively.
}
\caption{\label{del_rc}
Comparison of the results for the relative errors,
$\Delta r_{ch}= (r_{ch}^{expt}-r_{ch}^{th})/r_{ch}^{expt}$, in the values for the charge
rms radii for
nuclei considered in the fit to obtain the
 GSkI, GSkII and SSk forces.
The rms error for the charge rms radii are 0.023, 0.026 and 0.021 fm
for the GSkI, GSkII and SSk forces, respectively.
}
\caption{\label{pb208_spe}
The single particle energies for $^{208}Pb$ obtained using GSkI, GSkII and
SSk forces are compared with the experimental data for (a)
neutrons and 
(b) protons. The rms error in the single particle energies are 1.44, 1.13
and 1.67 MeV for the GSkI, GSkII and SSk forces, respectively. 
}
\caption{\label{nskin}
 Comparison of neutron skin thikness, $r_{n}-r_{p}$ for the GSkI, GSkII
and SSk forces. The values for neutron skin thikness
for $^{132}$Sn and $^{208}$Pb nuclei represented by filled
circles are the recent prediction based on isospin diffusin data \cite{Chen05}.
A=48 represents the result for $^{48}Ca$ nucleus. The
result for $^{48}Ni$ is discussed in the text. 
}
\caption{\label{isgmr}
Variations of the strength function with excitation energy for the 
(a)isoscalar giant monopole resonance, (b)isoscalar giant dipole resonance
and (c)isoscalar giant quadrupole resonance. The results are obtained
using GSkI, GSkII and SSk forces for the  $^{208}$Pb nucleus.  }

\caption{\label{Yj} Variations of the proton and electron fractions with
the baryon density for the GSkI and SSk forces.}
\end{figure}
\begin{figure}

\caption{\label{mrho}   The neutron star mass as a function of the
central baryon density $\rho_c$ for the GSkI and SSk forces.} 
\caption{\label{mrns} Relation between the neutron star mass and
its radius $R$ for the GSkI and SSk forces.}

\caption{\label{rp4}  Variation of $R P^{-1/4}$ as a function of
the baryon density $\rho$. Here, $R$ is the radius (in km) for the
neutron star of $1.4 M_\odot$ and $P$ is the pressure in units of
MeV/fm$^3$. The solid circles represent the empirical values taken
from Ref.   \cite{Lattimer05}.}

\caption{\label{ebind_beta} Ratio of the neutron star binding energy $E_{\rm
bind}$ to its corresponding gravitational mass ${ M}$ as a function
of the compactness parameter $\beta = \frac{G{ M}}{Rc^2}$. The solid
circles represent the empirical values calculated using  Eq. (\ref{ebind1}).}

\caption{\label{mi} Moment of inertia $I$ as a function of  the neutron star
mass. The values of $I$ are obtained at the Keplerian frequencies.}

\end{figure}

\begin{table}[p]
\caption{\label{range_skm} The lower (${\bf v_0}$) and upper (${\bf
v_1}$)  limits, maximum displacement (${\bf d}$) and initial values
(${\bf v}_{in}$)  for the Skyrme parameters used to minimize the $\chi^2$
value within the SAM.}

\begin{ruledtabular}
\begin{tabular}{|cdddd|}
\multicolumn{1}{|c}{}&
\multicolumn{1}{c}{${\bf v}_0$}&
\multicolumn{1}{c}{${\bf v}_1$}&
\multicolumn{1}{c}{${\bf d}$}&
\multicolumn{1}{c|}{${\bf v}_{in}$}\\
\hline
$t_0$(MeV$\cdot$fm$^3$)  & -3000.0 &  -1500.0 &  50.0& -1603.0\\
$t_1$(MeV$\cdot$fm$^5$)&-500.0&  500.0 & 20.0&515.9\\
$t_2$(MeV$\cdot$fm$^5$)  &-500.0&  500.0 & 20.0&84.5\\
$t_{31}$(MeV$\cdot$fm$^{3(\alpha_1+1)}$) &1000.0  & 3000.0&  50.0&1333.3\\
$t_{32}$ (MeV$\cdot$fm$^{3(\alpha_2+1)}$) & -1000  & 0.0 & 50.0&0.0\\
$t_{33}$ (MeV$\cdot$fm$^{3(\alpha_3+1)}$) & -500.0&  500.0 & 20.0&0.0\\
$x_0$&-4.0  &4.0 & 0.1&-0.02\\
$x_1$  &-4.0  &4.0 & 0.1&-0.5\\
$x_2$  &-4.0  &4.0 & 0.1&-1.713\\
$x_{31}$  & -4.0  &4.0 & 0.1& 0.1381\\
$x_{32}$  &-4.0  &4.0 & 0.1&0.0\\
$x_{33}$  &-4.0  &4.0 & 0.1&0.0\\
$\alpha_1$ & \frac{1}{3}&\frac{1}{3}&0 &\frac{1}{3}\\
$\alpha_2$ & \frac{2}{3}&\frac{2}{3}&0 &\frac{2}{3}\\
$\alpha_3$ & 1&1&0 &1\\
$W_0$(MeV$\cdot$fm$^5$)  &100.0 & 200.0 & 5.0&125.0\\
$V_w$ (MeV) &-3.0& 0.0 & 0.2& -2.05\\
$V_{w^\prime}$ (MeV) &0.0& 2.0 & 0.1& 0.697\\
$\lambda$  &400.0&  600.0&  10.0&485.0\\
$A_0$  &10.0&  50.0& 1.0&28.0\\
\end{tabular}
\end{ruledtabular}
\end{table}

\newpage
\begin{table}[p]
\caption{ \label{in2} 
Selected  experimental  data  for the binding energies $B$, charge rms
radii $r_{ch}$,  rms radii of valence neutron orbits $r_v$,
single-particle energies (S-P), breathing mode constrained
energies $E_0$ and EOS for the pure neutron matter used in the  fit to
determine the parameters of the Skyrme interaction.}
\begin{ruledtabular}
\begin{tabular}{|ccc|}
Properties& Nuclei& Ref.\\
\hline
$B$ & $^{16, 24}$O, $^{40,48}$Ca, $^{48,56,68,78}$Ni,
 $^{88}$Sr, $^{90}$Zr, $^{100,132}$Sn, $^{208}$Pb& \cite{Audi03}\\
$r_{ch}$& $^{16}$O, $^{40,48}$Ca, $^{56}$Ni, $^{88}$Sr, $^{90}$Zr,
$^{208}$Pb& \cite{Otten89,Vries87}\\
$r_v(\nu 1d_{5/2})$&$^{17}$O  & \cite{Kalantar88}\\
$r_v(\nu 1f_{7/2})$&$^{41}$Ca & \cite{Platchkov88}\\
S-P energies &$^{208}$Pb&\cite{Bohr69,BrownP}\\
$E_{o}$& $^{90}$Zr and $^{208}$Pb & \cite{Youngblood99} \\
EOS &  pure neutron matter &  \cite{Wiringa88}\\
\end{tabular}
\end{ruledtabular}
\end{table}
\newpage
\begin{table}[p]
\caption{\label{newpar} The values of the Skyrme parameters for GSkI, GSkII
and SSk interactions obtained by minimizing the $\chi^2$.  }

\begin{ruledtabular}
\begin{tabular}{|cddd|}
\multicolumn{1}{|c}{}&
\multicolumn{1}{c}{GSkI}&
\multicolumn{1}{c}{GSkII}&
\multicolumn{1}{c|}{SSk}\\
\hline
$t_0$(MeV$\cdot$fm$^3$)  & -1855.45 &  -1855.99 &-2523.52 \\
$t_1$(MeV$\cdot$fm$^5$)&397.23& 393.08& 435.00\\
$t_2$(MeV$\cdot$fm$^5$)  &264.63& 266.08& -382.04\\
$t_{31}$(MeV$\cdot$fm$^{3(\alpha_1+1)}$) &2309.67&2307.15  & 2372.49\\
$t_{32}$ (MeV$\cdot$fm$^{3(\alpha_2+1)}$) &-449.01&
-448.28& --\\
$t_{33}$ (MeV$\cdot$fm$^{3(\alpha_3+1)}$) &-53.31&-- &-- \\
$x_0$&0.1180& 0.0909& 0.6835\\
$x_1$  &-1.7586& -0.7203&  -0.4519\\
$x_2$  & -1.8068& -1.8369& -0.9214\\
$x_{31}$  & 0.1261& -0.1005& 1.0508\\
$x_{32}$  &-1.1881& -0.3529& --\\
$x_{33}$  & -0.4594& --& --\\
$\alpha_1$ & \frac{1}{3}&\frac{1}{3}&0.1682\\
$\alpha_2$ & \frac{2}{3}&\frac{2}{3}&--\\
$\alpha_3$ &1&--&--\\
$W_0$(MeV$\cdot$fm$^5$)  & 169.57& 152.28& 131.98\\
$V_w$ (MeV) & -2.9944& -2.6683& -1.2124\\
$V_{w^\prime}$ (MeV) &0.7059& 0.6191& 0.3077\\
$\lambda$  & 538.23& 442.56& 461.43\\
$A_0$  &49.91& 47.24& 19.33\\
\end{tabular}
\end{ruledtabular}
\end{table}

\begin{table}[p]
\caption{\label{nm_par} Nuclear matter properties for the
GSkI, GSkII and SSk  interactions at the minimum value of  $\chi^2 $.
The quantities given below are: $B/A$ the binding energy per nucleon,
$K_\infty$ the nuclear matter incompressibility coefficient,
$J$ the symmetry energy, $L=3\rho\frac{dJ}{d\rho}$
related to the slope of the symmetry energy, $m^*/m$ is the ratio of the
isoscalar effective nucleon mass to the bare nucleon mass and  $\rho_s$
the saturation density.  Values for these quantities are obtained at
the saturation density.}

\begin{ruledtabular}
\begin{tabular}{|cddd|}
\multicolumn{1}{|c}{Parameter}&
\multicolumn{1}{c}{GSkI}&
\multicolumn{1}{c}{GSkII}&
\multicolumn{1}{c|}{SSk}\\
\hline
  $B/A$ (MeV) & 16.02  & 16.13  & 16.15\\
  $K_\infty$(MeV)&230.20  & 233.60   &229.17\\
  $J $ (MeV) &  32.03  &  34.15  &33.49\\
  $L$ (MeV) &  63.46  & 66.82  &52.75\\
  $m^*/m $  &   0.78  &   0.79 &0.72\\
  $\rho_{s}$  &   0.159  &   0.159  &0.161\\
\end{tabular}
\end{ruledtabular}
\end{table}

\begin{table}[p]
\caption{\label{extra_data}
The rms radii of the valence neutron orbits $r_v$ (fm) and  the breathing
mode constrained energies (MeV).  The experimental values ( and the
theoretical error $\sigma$) used in the fit to determine the Skyrme
parameters are taken as follows: the values of $r_v$ were  taken from
Ref. \cite{Kalantar88,Platchkov88} ($\sigma=0.06$ fm) and values of the
breathing mode constrained energies were taken from \cite{Youngblood99}
( $\sigma=0.5$ MeV for the  $^{90}$Zr nucleus and $\sigma= 0.3$ MeV for
$^{208}$Pb nucleus). The superscript $`a`$ denotes that the corresponding data
were not included in the fit.} 

\begin{ruledtabular}
\begin{tabular}{|ccdddd|}
\multicolumn{1}{|c}{}&
\multicolumn{1}{c}{}&
\multicolumn{1}{c}{Exp.}&
\multicolumn{1}{c}{GSkI}&
\multicolumn{1}{c}{GSkII}&
\multicolumn{1}{c|}{SSk}\\
\hline
$r_v$& $(\nu 1d_{5/2}) $ &3.36 & 3.35& 3.37& 3.34\\
&$ (\nu 1f_{7/2}) $ & 3.99 &4.01& 4.02& 4.05\\
\hline
$E_0$&$^{90}$Zr& 17.81& 18.12& 18.17& 18.03\\
&$^{116}$Sn$^a$& 15.90& 16.52& 16.59&16.43\\
&$^{144}$Sm$^a$& 15.25& 15.62& 15.67& 15.52\\
&$^{208}$Pb& 14.18& 13.69& 13.70& 13.66\\
\hline
\end{tabular}
\end{ruledtabular}
\end{table}

\newpage

\begin{table}[p]
\caption{\label{ecen}
Values of the centroid energy $\left(\frac{m_1}{m_0} \right)$,
constrained energy $\sqrt{\frac{m_1}{m_{-1}}}$ and scaling energy
$\sqrt{\frac{m_3}{m_{1}}}$ obtained using strength function calculated
within the RPA frame work. 
The experimental data given here are taken from Refs.
\cite{Youngblood99,Youngblood04,Youngblood04a}.}
\begin{ruledtabular}
\begin{tabular}{|ccdddd|}
\multicolumn{1}{|c}{Nuclei}&
\multicolumn{1}{c}{Interaction}&
\multicolumn{1}{c}{$\frac{m_1}{m_0}$}&
\multicolumn{1}{c}{$\sqrt{\frac{m_1}{m_{-1}}}$}&
\multicolumn{1}{c}{$\sqrt{\frac{m_3}{m_{1}}}$}&
\multicolumn{1}{c|}{$\Gamma$}\\
\hline
& & ISGMR&&&\\
\hline
$^{90}$Zr& GSkI&18.48&    18.25&    19.14&     2.70\\
& GSkII& 18.60&    18.36&    19.25&     2.72\\
   &SSk&18.43&    18.20&    19.05&     2.63\\
   &Exp. &17.89& 17.81& &  \\
$^{208}$Pb& GSkI&13.79&    13.55&    14.46&     2.33\\
   &GSkII &13.79&    13.55&    14.46&     2.33 \\
   &SSk&13.87&    13.64&    14.53&     2.31\\
   &Exp.& 14.17&    14.18 && \\
\hline
& & ISGDR&&&\\
\hline
$^{90}$Zr& GSkI&28.56&    28.11&    29.74&     4.82\\
& GSkII& 28.48&    28.03&    29.69&     4.87\\
   &SSk&28.83&    28.38&    30.04&     4.90\\
   &Exp.&26.7  & &&\\
$^{208}$Pb& GSkI&24.14&    23.87&    24.95&     3.60\\
   &GSkII&23.78&    23.50&    24.65&     3.69\\
   &SSk&24.61&    24.32&    25.46&     3.73\\
   &Exp. &22.20  & && \\
\hline
 & & ISGQR&&&\\
\hline
$^{90}$Zr& GSkI& 14.65&    14.50&    15.52&     2.65\\
& GSkII&  14.42&    14.29&    15.22&     2.50\\
   &SSk&15.10&    14.98&    15.86&     2.50\\
   &Exp. & 14.30&  & &\\
$^{208}$Pb& GSkI&11.34&    11.23&    12.10&     2.10\\
   &GSkII& 11.07&    10.97&    11.82&     2.05\\
   &SSk&11.92&    11.80&    12.71&     2.23\\
   &Exp. &10.89  && & \\
\hline
\end{tabular}
\end{ruledtabular}
\end{table}
\begin{table}[p]
\caption{\label{ns_prop}
Some of the  key properties of the neutron star with $1.4 M_\odot$ and the
maximum mass $M_{\rm max}$ calculated  using the  GSkI and SSk forces.}
\begin{ruledtabular}
\begin{tabular}{|cdddd|}
\multicolumn{1}{|c}{}&
\multicolumn{2}{c}{$1.4 M_\odot$}&
\multicolumn{2}{c|}{$M_{\rm max}$}\\
\cline{2-5}
 & GSkI & SSk& GSkI& SSk\\
\hline
$\rho_c$ (fm$^{-3}$)& 0.53&0.55&1.21&1.22\\
$\varepsilon_c/c^2$ ($10^{14}$ g cm$^{-3}$)& 9.68& 10.03&27.97&  28.54\\
$M (M_\odot)$& 1.40& 1.40& 1.95& 2.00\\
R (km)& 11.97& 11.62& 10.05& 9.87\\
$A$ ($10^{57}$)& 1.81& 1.82& 2.68& 2.77\\
$E_{\rm bind}$ ($10^{53}$ erg)& 2.03& 2.09& 5.09& 5.59\\
$z_{\rm surf}$ &0.24 & 0.25 & 0.53& 0.58\\
\hline
\end{tabular}
\end{ruledtabular}
\end{table}
\include{fig}
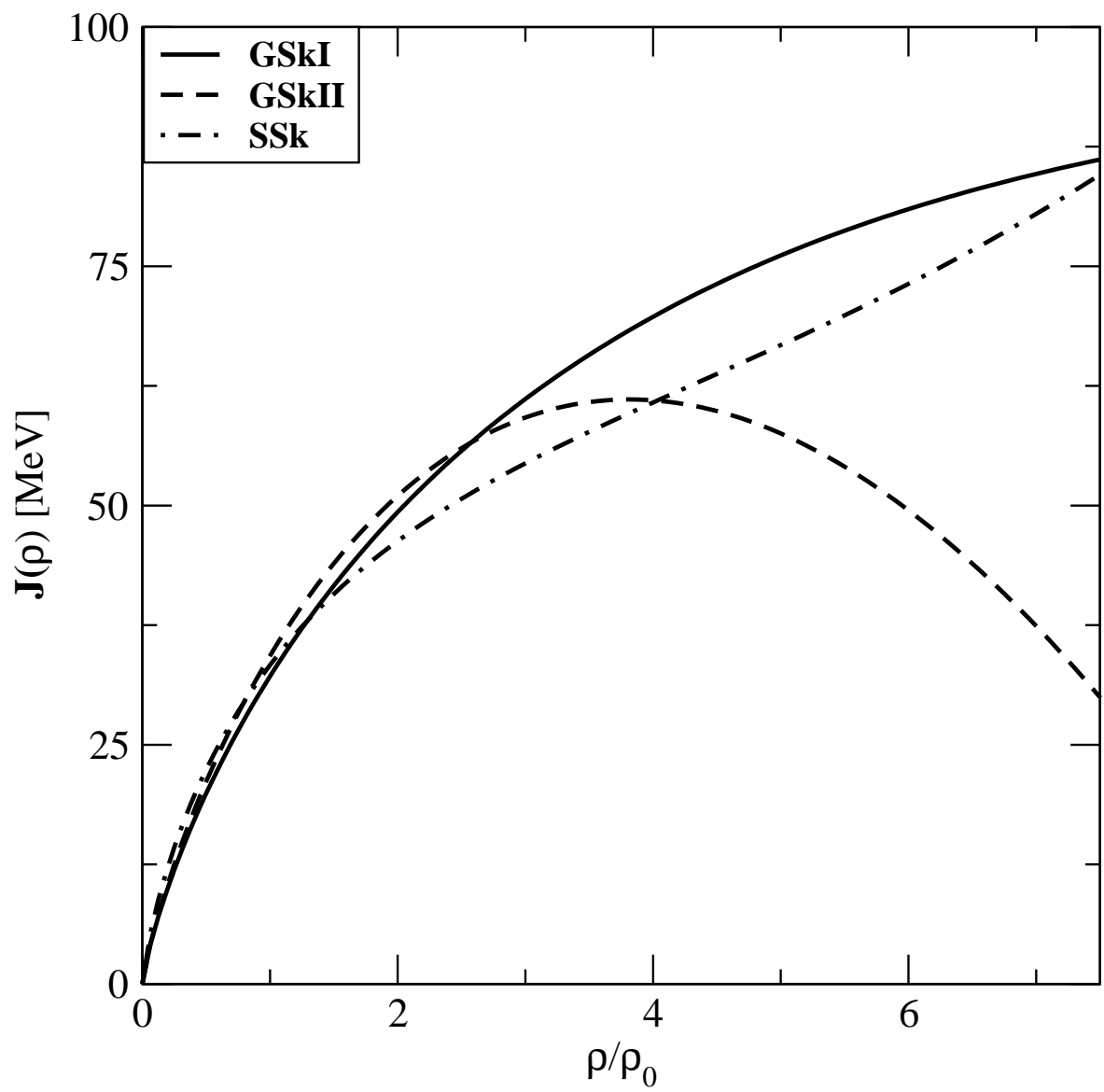
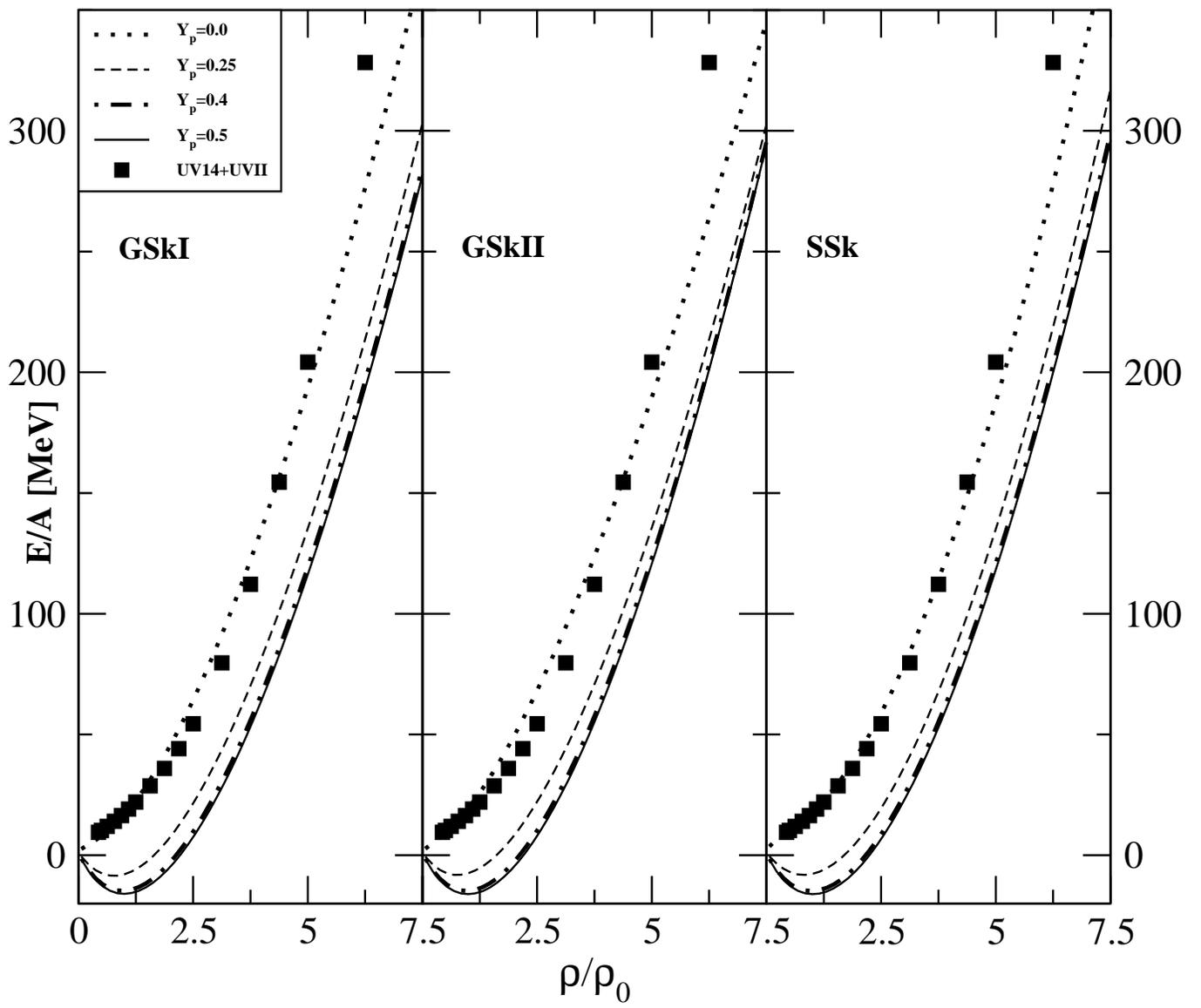
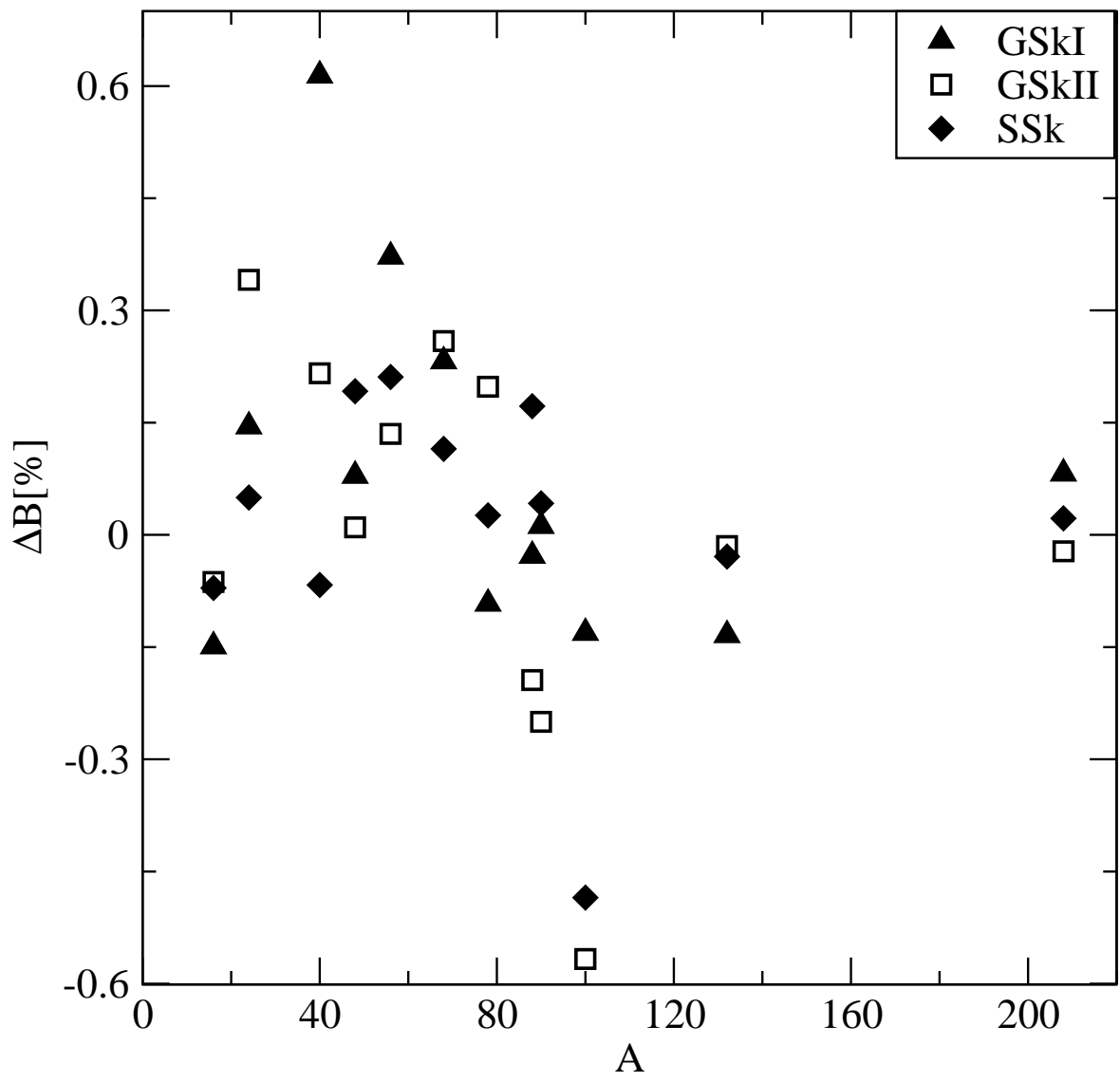
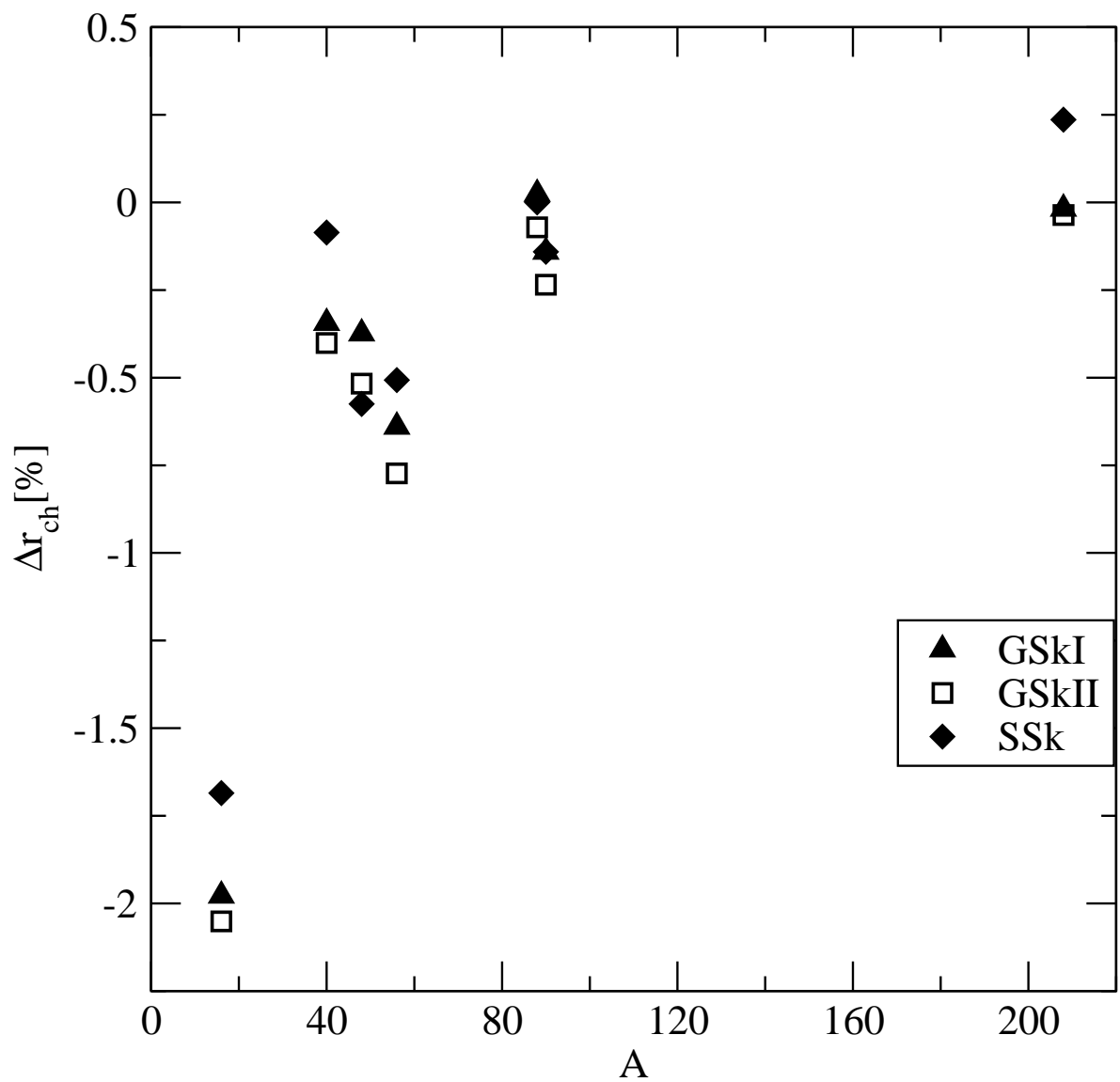
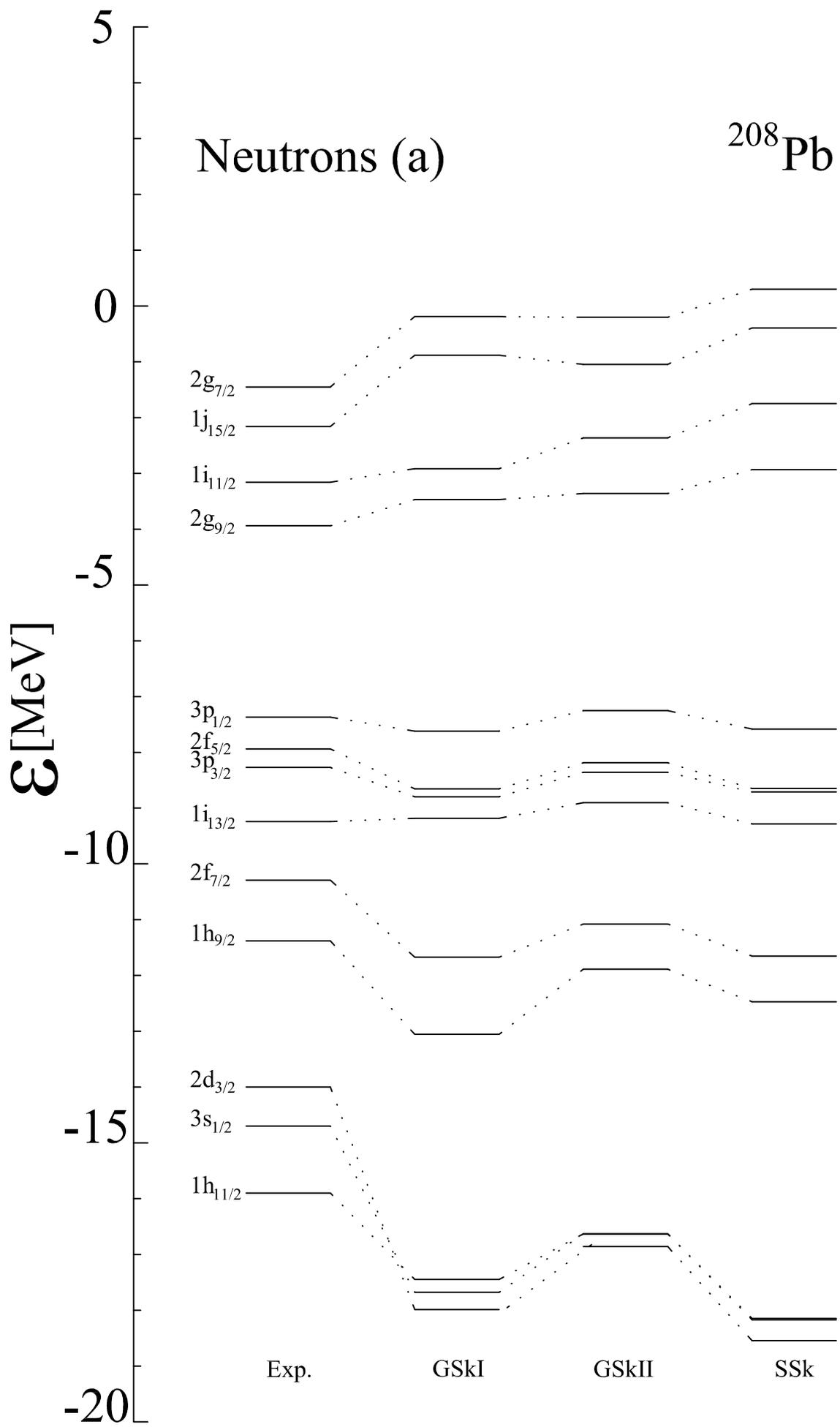
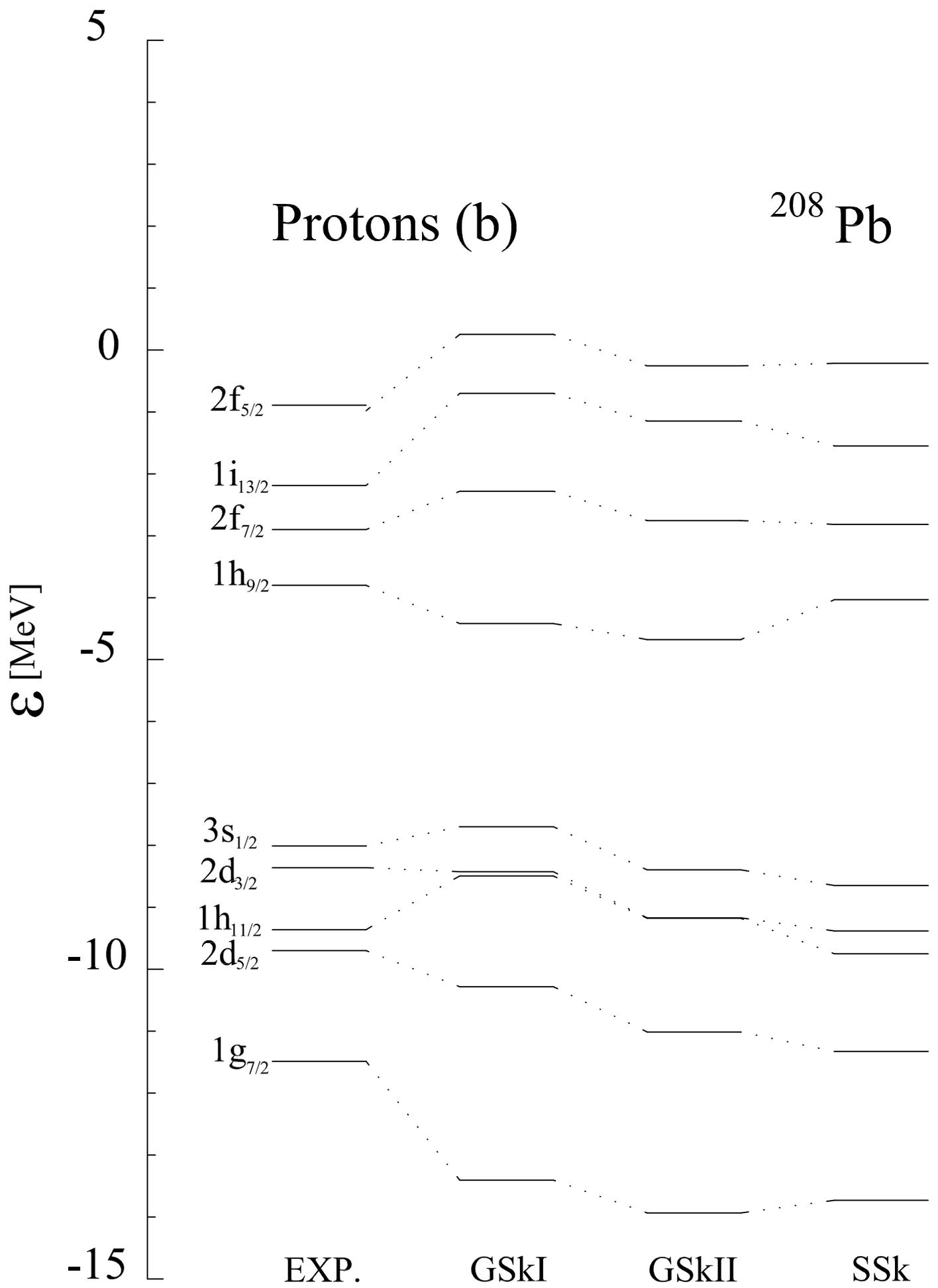
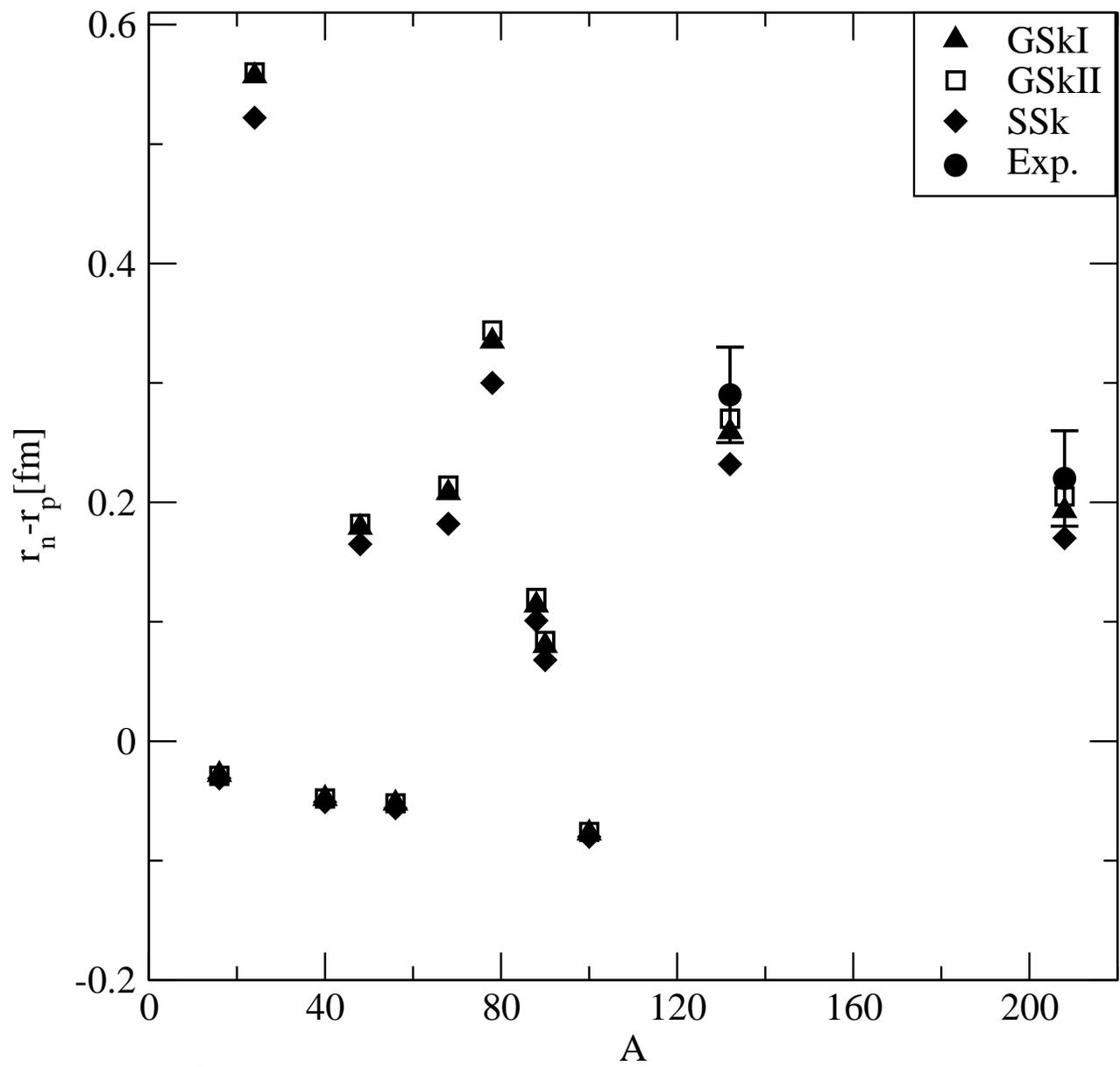
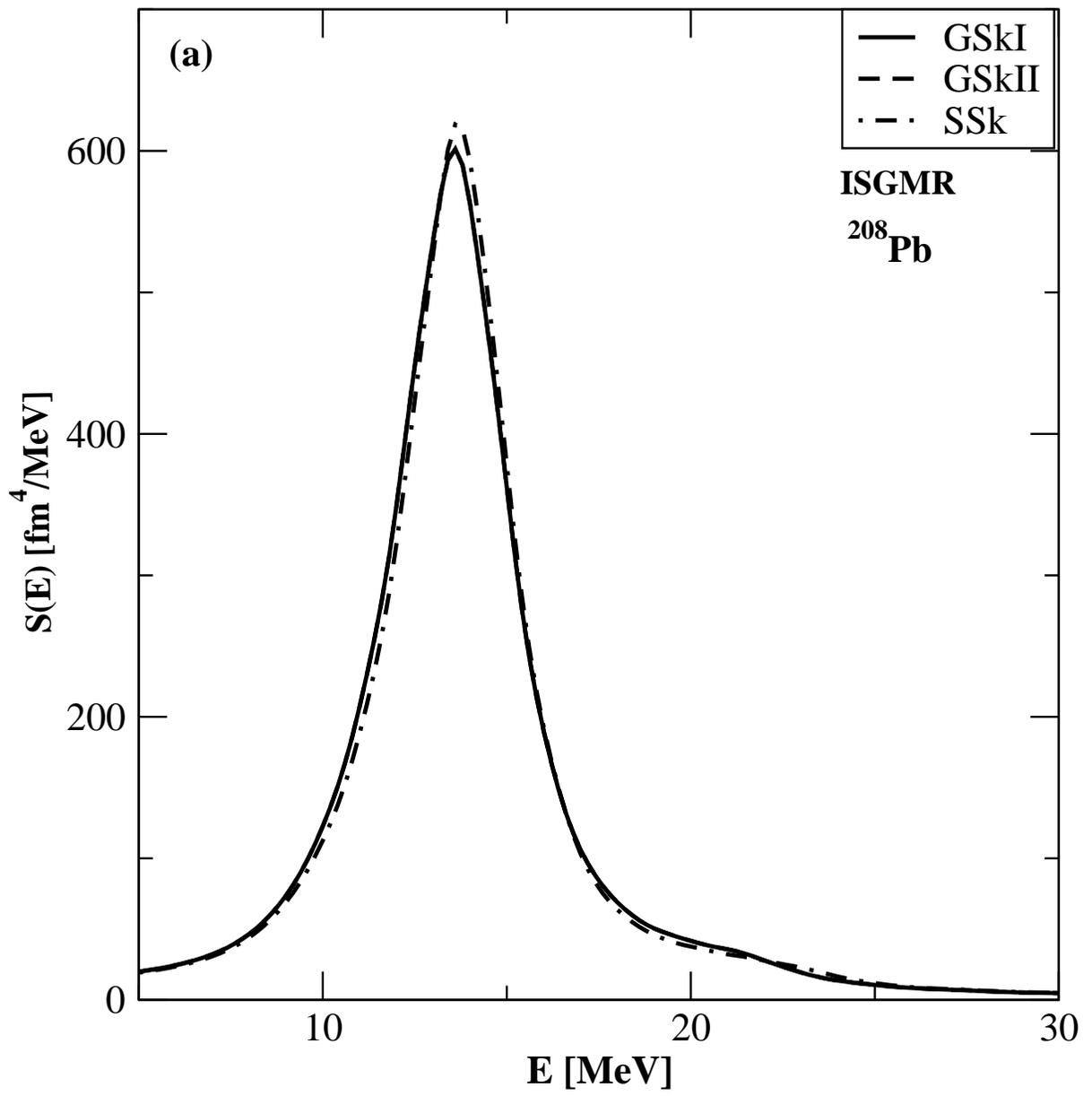
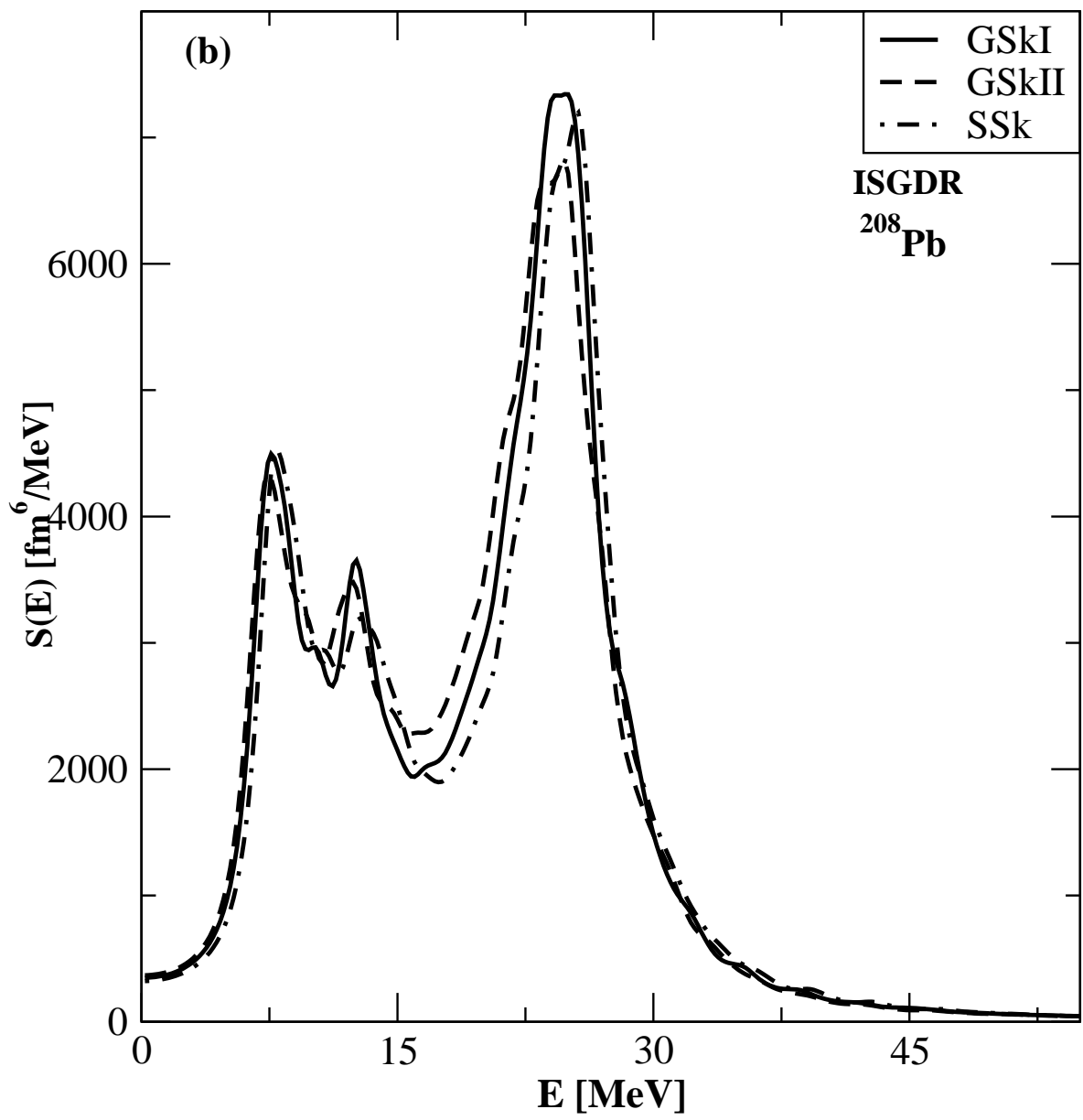
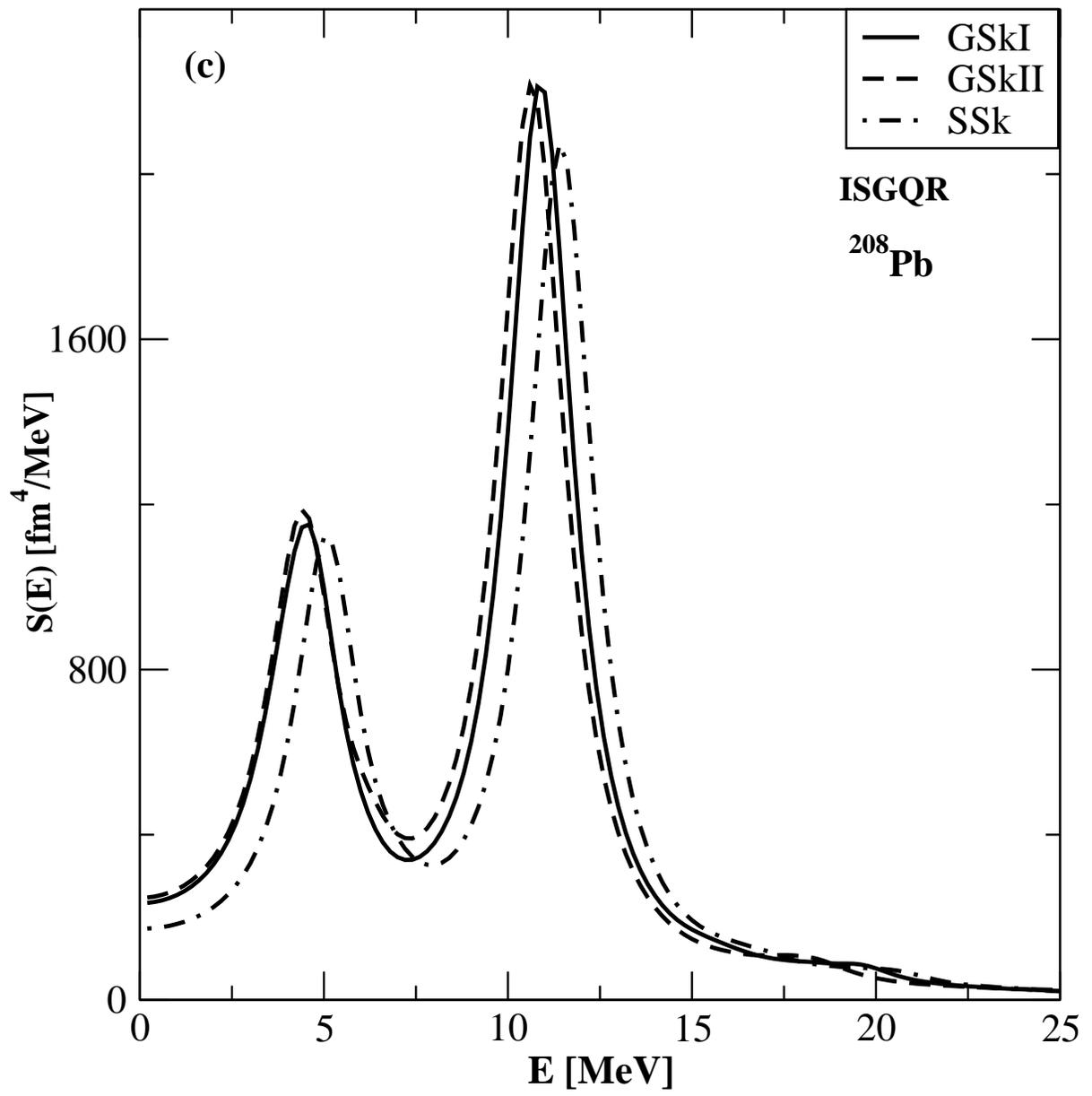
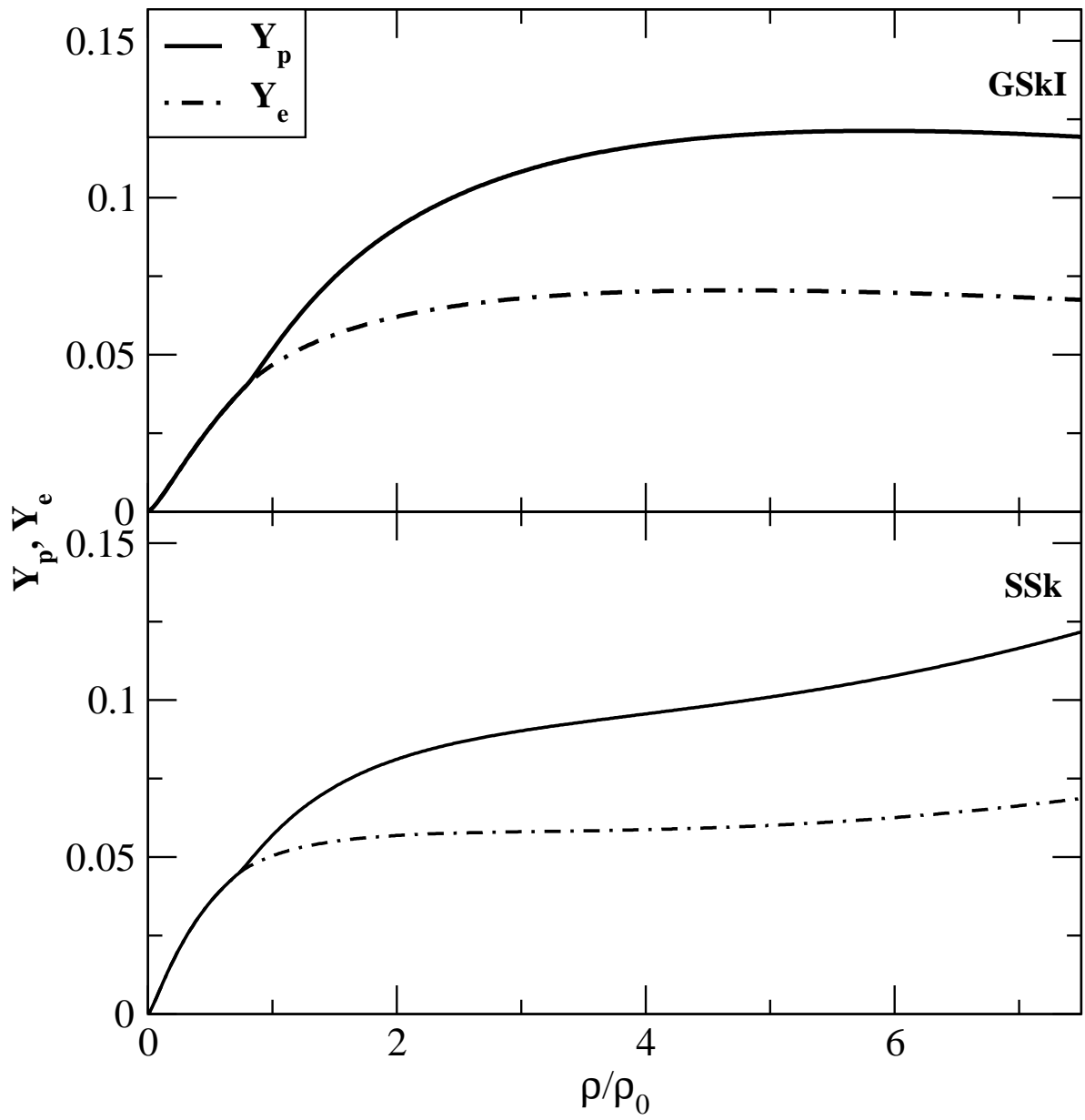
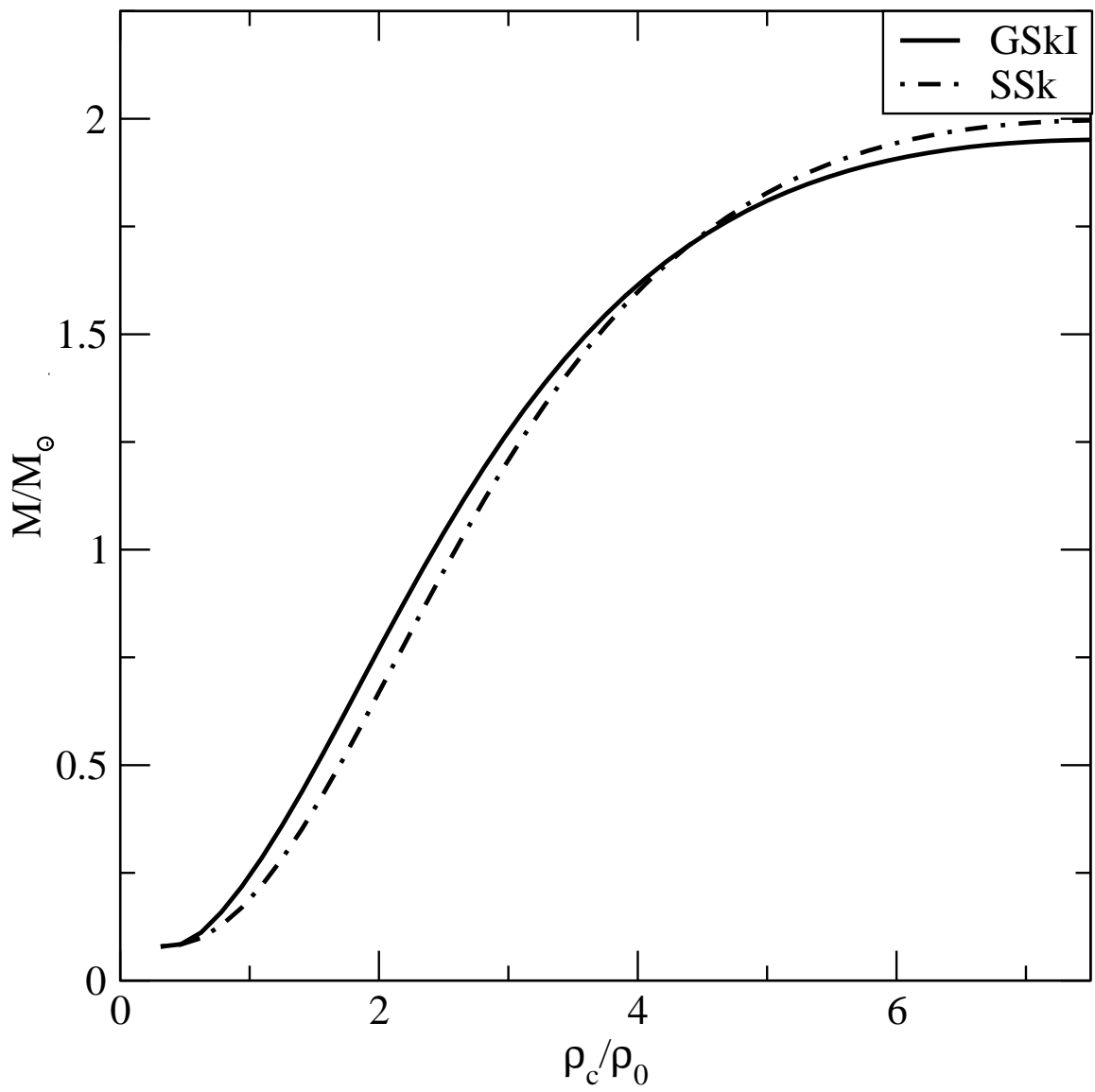
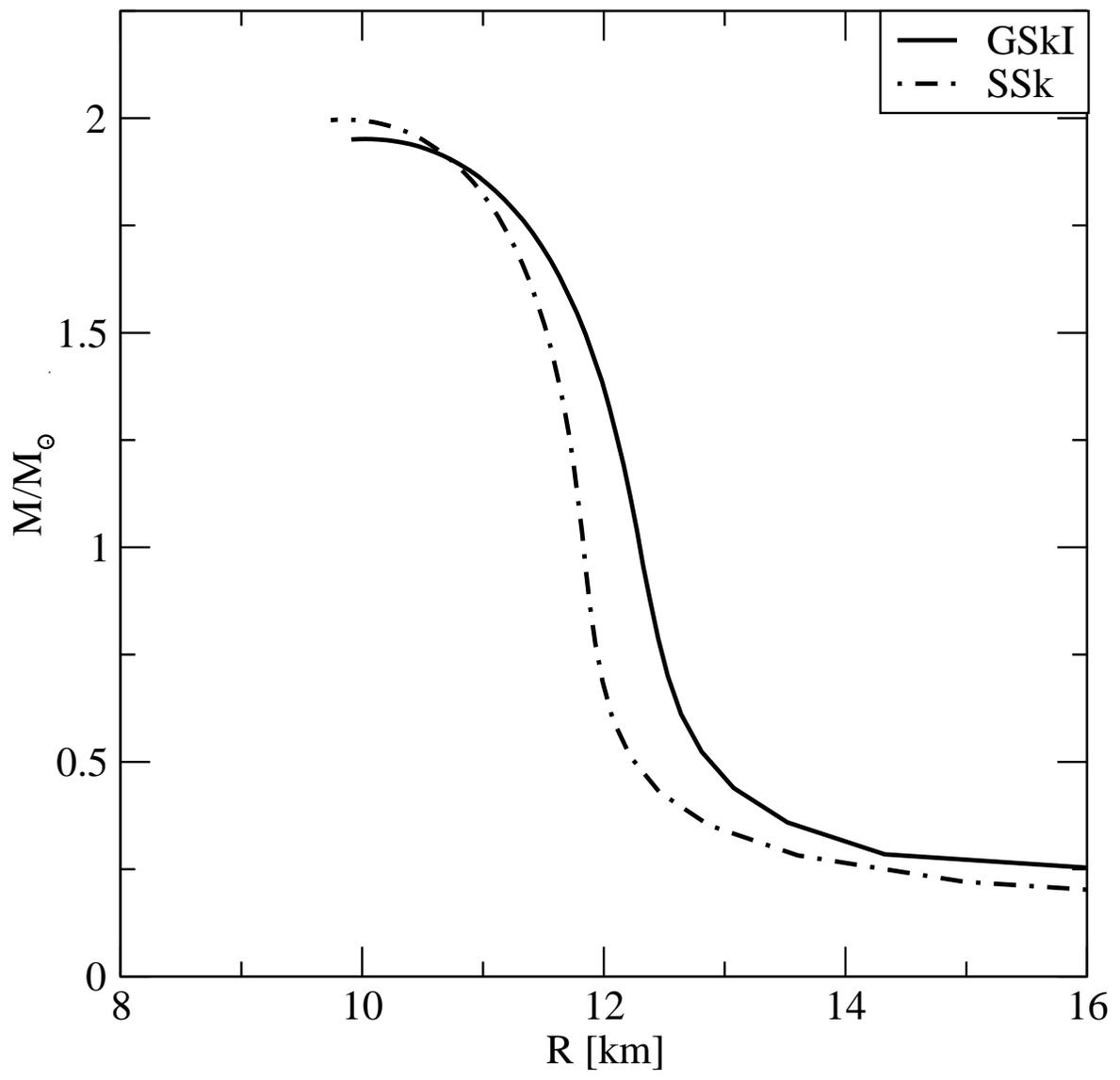
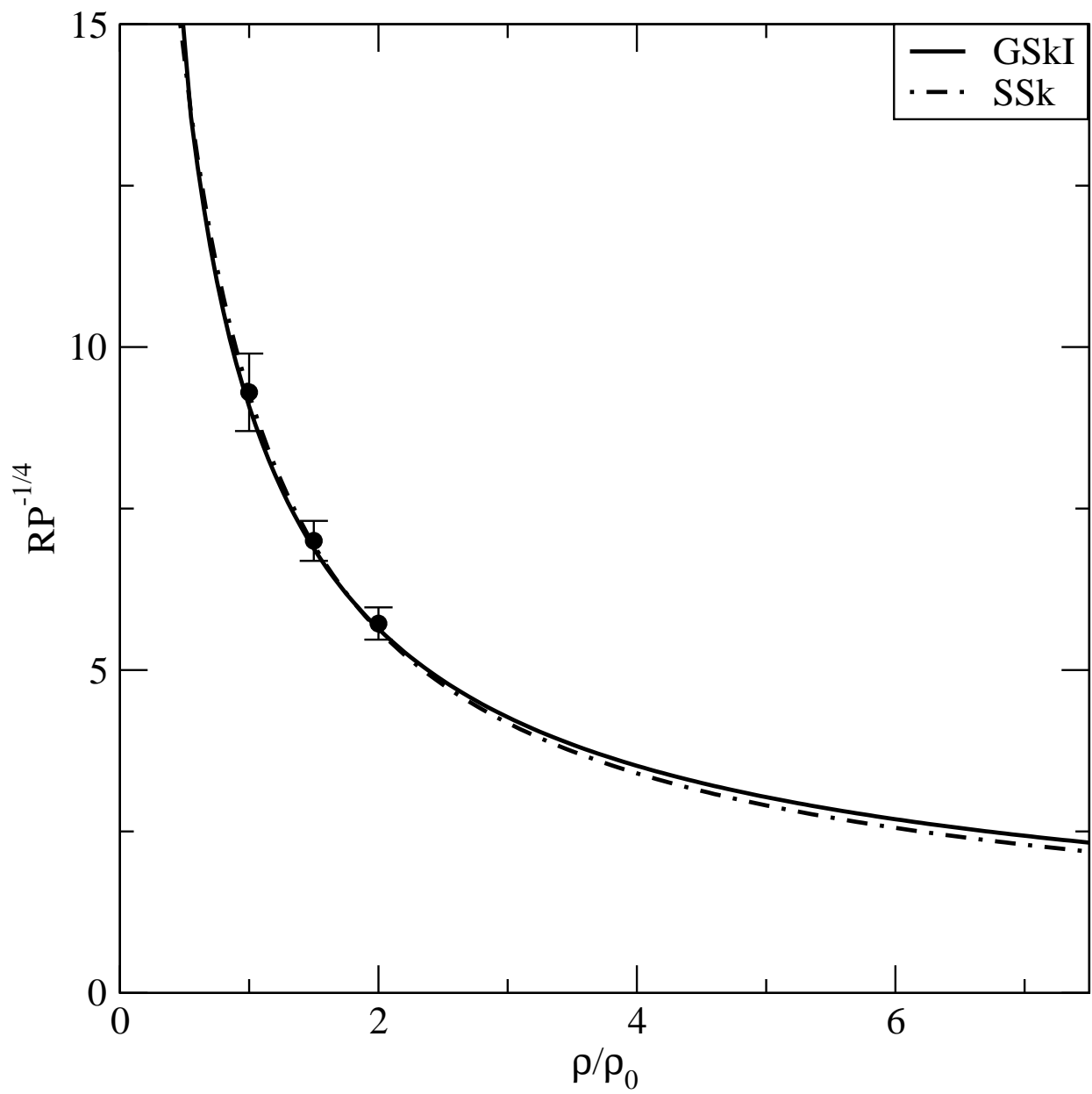
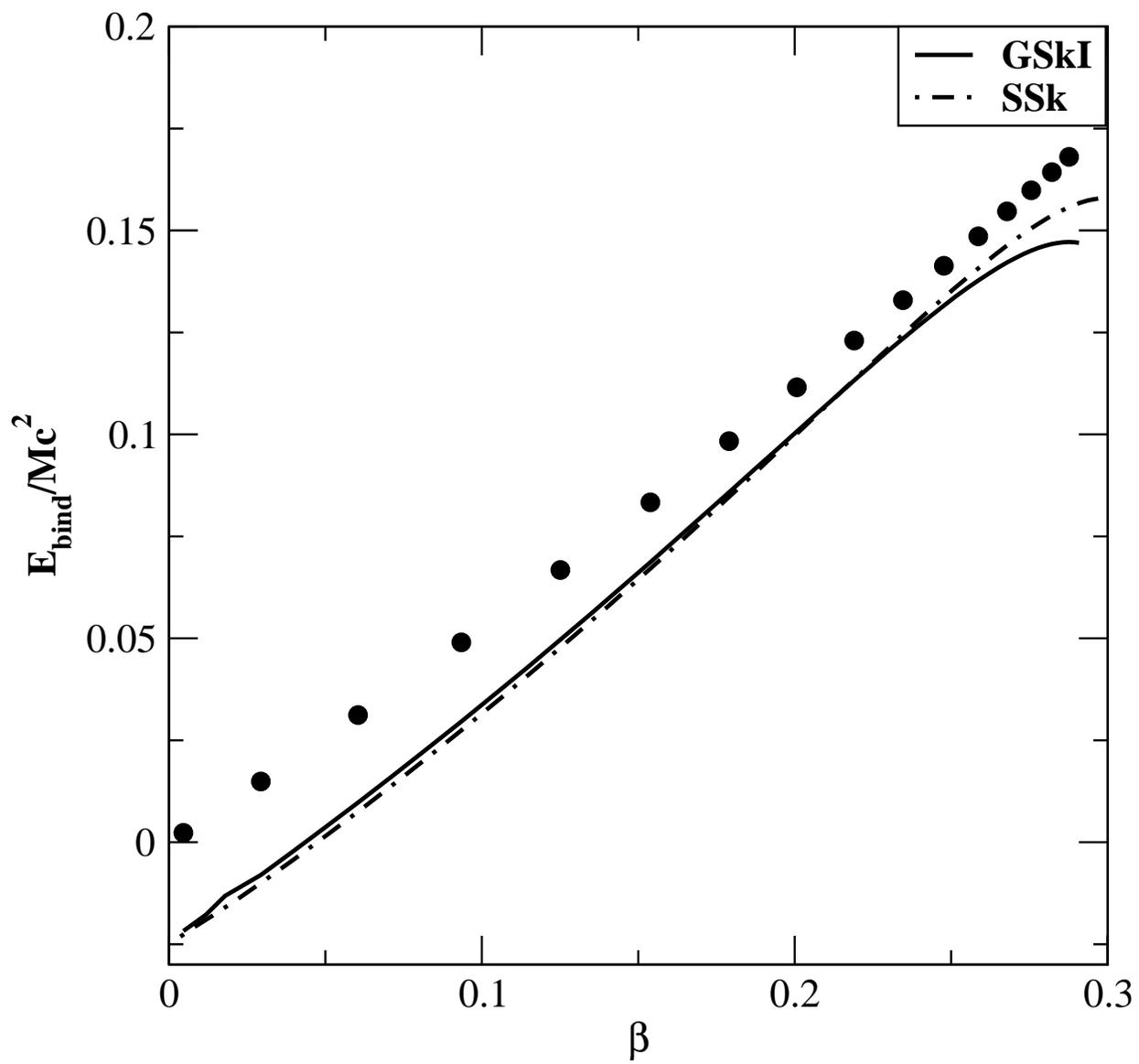
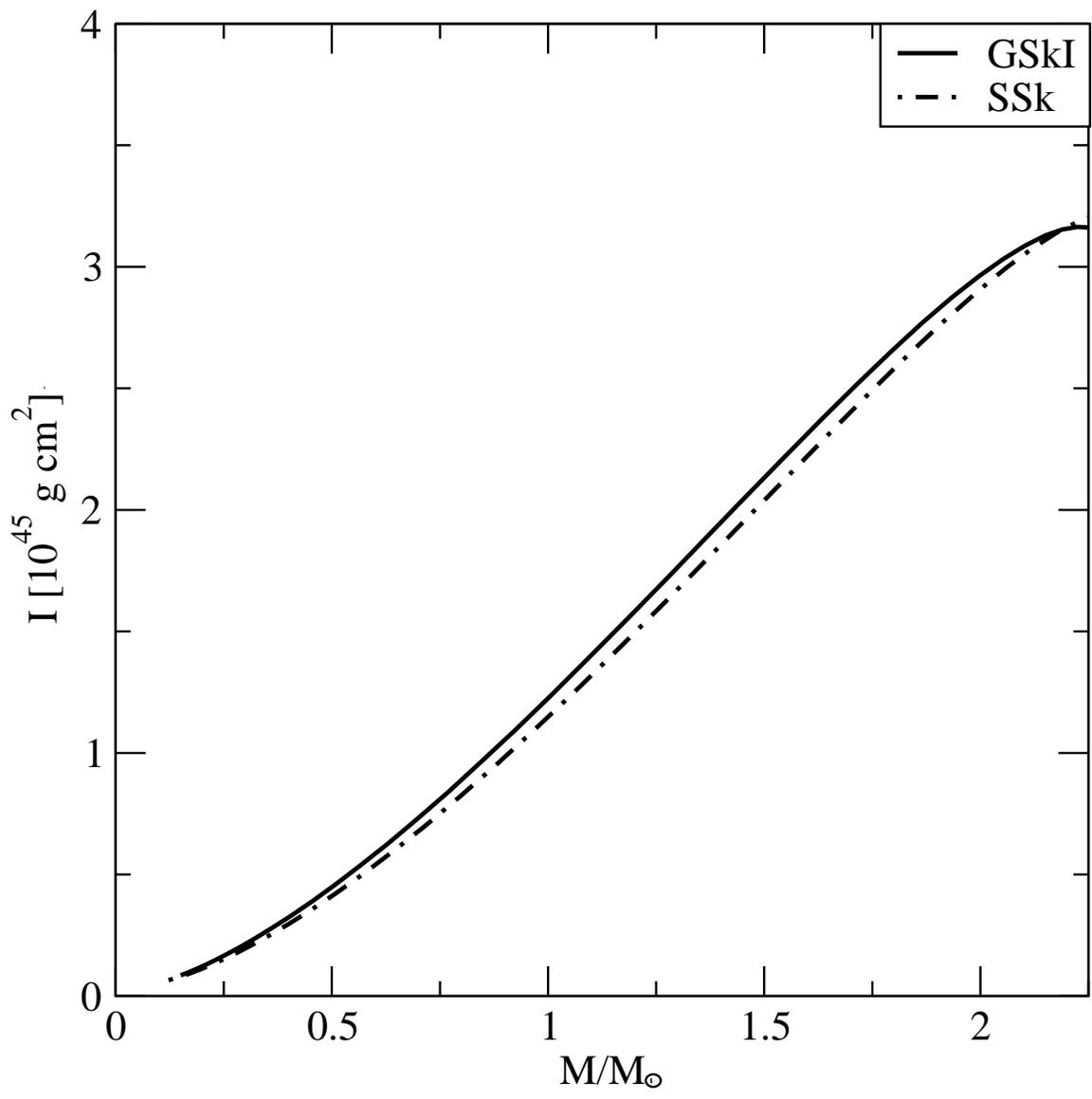
\end{document}